\documentclass[12pt,preprint]{aastex}

\usepackage{emulateapj5}
\newcommand{\et}{et al.}

\newcommand{\xte}{{\it RXTE}}

\newcommand{\xmm}{{\it XMM-Newton}}

\newcommand{\asca}{{\it ASCA}}

\newcommand{\Msun}{\hbox{$\rm\thinspace M_{\odot}$}}
\slugcomment{}
\shortauthors{Markowitz et al.}
\shorttitle{Fe K Emission and Absorption in IC~4329a}

\begin{document}
\title{Fe K Emission and Absorption in the XMM-EPIC Spectrum of the Seyfert Galaxy IC~4329a} 

\author{A.~Markowitz\altaffilmark{1,2}, J.N.~Reeves\altaffilmark{1,3}, V.~Braito\altaffilmark{1,3}
\altaffiltext{1}{X-Ray Astrophysics Laboratory, Code 662, NASA Goddard Space Flight Center, Greenbelt, MD 20771; agm, jnr, vale@milkyway.gsfc.nasa.gov}
\altaffiltext{2}{NASA Post-doc Research Associate}
\altaffiltext{3}{Department of Physics and Astronomy, Johns Hopkins
University, Baltimore, MD 21218}
}

\begin{abstract}

We present a detailed analysis of the {\it XMM-Newton} long-look of the 
Seyfert galaxy IC~4329a. The Fe K bandpass is 
dominated by two resolved peaks at 6.4 keV and 7.0 keV, consistent with 
neutral or near-neutral Fe K$\alpha$ and K$\beta$ emission.
There is a prominent redward asymmetry in the 6.4 keV line,
which could indicate emission from a Compton shoulder.
Alternatively, models using dual relativistic disklines are 
found to describe the emission profile well. A low-inclination, 
moderately relativistic dual-diskline model is possible if the 
contribution from narrow components, due to 
distant material, is small or absent. A high-inclination, moderately 
relativistic profile for each peak is 
possible if there are roughly equal
contributions from both the broad and narrow components. 
Combining the {\it XMM-Newton} data with {\it RXTE} monitoring data, we
explore the time-resolved spectral behavior on time scales 
from hours to 2 years. We find no strong 
evidence for variability of the Fe K line flux on any time 
scale, likely due to the minimal level of continuum variability.
We detect, at high significance, a narrow absorption line at 7.68 keV.
This feature is most likely due to 
\ion{Fe}{26} K$\alpha$ absorption blueshifted to $\sim$0.1$c$ 
relative to the systemic velocity, suggesting a high-velocity, 
highly ionized outflow component. As is often the case with similar 
outflows seen in high-luminosity quasars, the power associated with the
outflow represents a substantial 
portion of the total energy budget of the AGN. 
The outflow could arise from a radiatively-driven disk wind, or it may be 
in the form of a discrete, transient blob of ejected material.
\end{abstract}


\section{Introduction}

The hard X-ray emission of Seyfert 1 AGN is dominated by rapidly-variable 
emission thought to originate from inverse Comptonization of soft seed photons 
by a hot corona near the central black hole (e.g., Shapiro, Eardley \& 
Lightman 1976; Sunyaev \& Titarchuk 1980; Haardt \et\ 1994). %%%Zdz2003???  
Furthermore, the disk, or some other cold, optically thick material, 
reprocesses the hard X-rays, as evidenced by the so-called 'Compton 
reflection humps' seen in Seyfert spectra above $\sim$7~keV and peaking 
near 20--30 keV (Pounds \et\ 1990).

Another signature of reflection is the Fe K$\alpha$ emission line commonly 
seen at 6.4 keV; this has proved to be a valuable diagnostic of the 
accreting material near black holes. A few sources show evidence for 
a relativistically broadened component that is likely produced in the 
inner accretion disk, its profile sculpted by gravitational redshifting
and relativistic Doppler effects (e.g., Fabian \et\ 2002). However, a 
narrow Fe K$\alpha$ component is a much more common feature in Seyfert 1s 
(e.g., O'Brien 2001, Yaqoob \et\ 2001). FWHMs of several thousand 
km s$^{-1}$ are typical (e.g., Yaqoob \& Padmanabhan 2004). The narrow 
component is generally thought to originate in Compton-thick material 
far from the central black hole, such as the outer accretion disk, 
the putative molecular torus invoked in standard unification models, or
Compton-thick gas commensurate with the Broad-Line Region (BLR).

At the same time, there is strong evidence from X-ray and UV grating 
observations for the presence of ionized material in the inner regions 
of a large fraction of AGN (e.g., Blustin \et\ 2005). High-resolution 
spectroscopy shows that the gas is usually outflowing from the nucleus; 
typical velocities are $\sim$ a few hundred km s$^{-1}$. Absorption due 
to a broad range of ionic species is commonly seen. For many sources, 
the relative line strengths argue for several different photoionized 
absorbing components, as opposed to a single absorber, along the line of 
sight. In the Fe K bandpass, absorption at 6.7 keV has been seen in 
NGC 3783 with \xmm\ (Reeves \et\ 2004), consistent with absorption by 
\ion{Fe}{25}. In addition, absorption features near $\sim$7--8 keV
have been detected in PG and BAL quasars(Reeves \et\ 2003;
Pounds \et\ 2003b; Chartas \et\ 2002) as well as in the
high-luminosity Seyfert galaxy Mkn 509 (Dadina \et\ 2005);
such features have been attributed to strongly blueshifted,
high-ionization Fe K-shell absorption at near-relativistic
($\sim$0.2$c$) velocities. These features may be signatures of high-velocity accretion
disk winds.

IC 4329a is a well-studied, nearby (z = 0.01605: Willmer \et\ 1991) 
Seyfert 1.2 nucleus
embedded in a nearly edge-on host galaxy. It has consistently remained 
one of the X-ray-brightest Seyfert galaxies for at least a decade. 
Using simultaneous \asca\ and \xte\ data, Done, Madejski \& 
$\dot{\rm Z}$ycki (2000) found the 6.4 keV Fe K$\alpha$ core to be 
moderately broadened ($\sim$20,000 km s$^{-1}$ FWHM). The emission was 
parameterized using a relativistic diskline model, with emission extending 
inward only to 30--100 gravitational radii $R_{\rm g}$. Using a 60 ksec 
{\it Chandra}-HETGS observation performed in 2001, McKernan \& Yaqoob 
(2004) not only detected a narrow 6.4 keV core, but also confirmed the 
presence of emission near 6.9 keV. This double-peaked complex could be 
fit by several competing models, included dual Gaussians, dual disklines, 
or a single diskline.  Steenbrugge \et\ (2005; hereafter S05) observed
IC~4329a with \xmm\ for 136 ksec in 2003 (see $\S$2 for details). Their 
analysis concentrated primarily on the soft X-ray, RGS data, which revealed 
evidence for several absorbing components, including neutral 
absorption intrinsic to IC~4329a's host galaxy, and a four-component 
warm absorber spanning a range of ionization states (the log of 
the ionization parameter $\xi$ ranged from --1.4 to +2.7). Evidence 
for absorption due to local, z=0 hot gas was also detected. The 
EPIC-pn data (see below) also detected two emission peaks near 
6.4 and 6.9 keV (rest frame), identified as the Fe K$\alpha$ core 
and a blend of \ion{Fe}{1} K$\beta$ and \ion{Fe}{26} emission, 
respectively. S05 reported that the 6.4 keV line was narrow 
and did not vary with time, consistent with an origin far from 
the central black hole.

In this paper, we aim not only to parameterize the Fe K emission due 
to accreting material in IC 4329a, but also to search for absorption 
features in the Fe K bandpass that could be indicative of outflowing 
material. To this end, we have re-analyzed the \xmm\ pn spectrum, 
concentrating on the Fe K bandpass. We also augment the \xmm\ data 
with spectra derived from \xte\ monitoring data to explore time-resolved 
spectroscopy over a wide range of time scales, search for variability 
in the Fe K core flux, and parameterize the relation between the core 
and the flux of the X-ray continuum. The \xmm\ and \xte\ observations 
and data reduction are described in $\S$2. Spectral fits to the \xte\ 
data are described in $\S$3. Spectral fits to the Fe K bandpass of 
the \xmm\ EPIC data are described in $\S$4. Time-resolved spectral 
fitting to the \xte\ and \xmm\ data are described in $\S$5. The results 
are discussed in $\S$6, followed by a brief summary in $\S$7.

\section{Observations and Data Reduction}
 
\subsection{XMM-EPIC}
 
IC~4329a was observed by \xmm\ during revolution 670 on 2003 Aug 
6--7, for a duration of 136 ksec. This paper uses data taken with the 
European Photon Imaging Camera (EPIC), which consists of one pn CCD 
back-illuminated array sensitive to 0.15--15 keV photons (Str\"{u}der 
\et\ 2001), and two MOS CCD front-illuminated arrays sensitive to 
0.15--12 keV photons (MOS1 and MOS2, Turner \et\ 2001). Data from 
the pn were taken in Small Window Mode, data from the MOS1 was in 
PrimePartialW3/large window mode, and data from the MOS2 was in 
PrimePartialW2/small window mode. The thin filters were used for 
the pn and MOS2 cameras; the MOS1 used the medium filter. Spectra 
were extracted using {\it XMM}-SAS version 6.50 and using standard 
extraction procedures. For all three cameras, source data were 
extracted from a circular region of radius 40$\arcsec$; backgrounds were
extracted from circles of identical size, centered $\sim$3$\arcmin$ away
from the core.  
Hot, flickering, or bad pixels were excluded. Data were
selected using event patterns 0--4 for the pn. To reduce pile-up, 
only pattern 0 events were extracted for the MOS cameras. However, 
we found (e.g., using the SAS task {\sc epatplot}) that the MOS1 data 
above 2 keV were severely piled-up; the level of pile-up ranged from 
$\sim$5--10$\%$ at 2--3 keV to $\sim$50$\%$ at 10 keV. We do not 
consider the MOS1 data further in this paper. The MOS2 data above 
6 keV suffered from a small level of pile-up, ranging from $\sim$3$\%$ 
at 6 keV to 10$\%$ at 10 keV.
%%%%%% We inspected the background light curves; due to increased 
%%%%%%% background levels at the end of the satellite orbit, data 
%%%%%%% after 2003 Aug 7 at 19:49 UTC were excluded; this did not 
%%%%%%% affect any MOS data but resulted in the exclusion of the 
%%%%%%% final 0.6 ksec of pn data. 
This reduction yielded 133.3 ksec of pn data, starting at
2003 Aug 6 at 06:57 UTC, and 132.9 ksec of MOS2 data,
starting at 2003 Aug 6 at 06:51 UTC. After correcting for deadtime 
effects, the final exposure times for the pn and MOS2 were 93.2 ksec 
and 128.8 ksec, 
respectively. This paper will concentrate 
primarily on data obtained with the higher signal-to-noise pn, although 
MOS2 data were also analyzed for consistency. 
Background-subtracted pn light curves have already been presented in 
S05. We found a mean 2--10 keV pn count rate of 6.4 c s$^{-1}$, 
corresponding to a flux of $8.9 \times 10^{-11}$ erg cm$^{-2}$ s$^{-1}$.
The pn and MOS2 spectra were grouped to a minimum of 50 counts per bin
with {\sc grppha}.

Since background flares due to proton flux tend to have hard spectra, 
the 10--12 keV band of the pn is the band most sensitive to them. The 10--12 
keV pn background light curve, plotted in log-space, is shown in Figure 1.
To test if these flares had any impact on the pn source spectrum,
we filtered out data collected during times when the 10--12 keV
background light curve exceeded a count rate of $B$ + 2$\sigma$ (equal to 
0.3 c s$^{-1}$), where $B$ is the mean background rate and $\sigma$ is the 
standard deviation of the light curve. This screening yielded a new 
exposure time of 71.6 ksec. We found that the only major difference was 
that the screened spectra tended to be fit with a photon index 
approximately 0.01 steeper than the unscreened spectra, but 
all other parameters remained virtually unchanged. In particular, the 
Fe K profile, the focus of this paper, remained unchanged. We therefore 
concentrated on the unscreened data in all analysis below to increase 
the signal-to-noise ratio, although we checked the screened data for consistency. 
However, we note that the brightness of the sources could have an impact 
on the degree of difference between the screened and unscreened spectra; 
IC 4329a is an X-ray bright AGN
(the 2.5--12 keV source/background count rate ratio is at least 25), 
and for fainter sources, such screening 
could potentially have a much larger impact than for this data set.

\subsection{\xte\ observations and data reduction}
 
IC~4329a has been monitored by \xte\ once every 4.3 d since 
2003 Apr 8 (hereafter referred to as long-term sampling).  In this paper, 
we include data taken up until 2005 Oct 02. Due to sun-angle constraints, 
no data were taken from 2003 Oct 03 -- 2003 Nov 20 or from 
2004 Oct 04 -- 2004 Nov 21. IC~4329a was subjected to intensive \xte\ 
monitoring, once every third orbit (17 ksec) from 2003 Jul 10 to 
2003 Aug 13 (hereafter referred to as medium-term sampling; short-term 
sampling refers to the \xmm\ long-look detailed above). Each \xte\ 
visit lasted $\sim$1 ksec.

\subsubsection{PCA Data Reduction}

Data were taken using {\it RXTE}'s proportional counter array 
(PCA; Swank 1998), which consists of five identical collimated
proportional counter units (PCUs). For simplicity, data were collected
only from PCU2 (PCU0 suffered a propane layer loss in 2000 May; the other 
three PCUs tend to suffer from repeated breakdown during on-source time).  
Count rates quoted in this paper are per 1 PCU. The data were reduced 
using standard extraction methods and {\sc FTOOLS~v5.3.1} software. 
Data were rejected if they were gathered less than 10$\arcdeg$ from 
the Earth's limb, if they were obtained within 30~min after the 
satellite's passage through the South Atlantic Anomaly, if 
{\sc ELECTRON2}~$>$~0.1, or if the satellite's pointing offset was greater
than 0$\fdg$02.

As the PCA has no simultaneous background monitoring capability, background
data were estimated by using {\sc pcabackest~v3.0} to generate model files
based on the particle-induced background, SAA activity, and the diffuse X-ray
background. The 'L7-240' background models appropriate for faint sources
were used. This background subtraction is the dominant source of systematic
error in \xte\ AGN monitoring data (e.g., Edelson \& Nandra 1999). Counts 
were extracted only from the topmost PCU layer to maximize the 
signal-to-noise ratio. Spectra from individual visits were merged to form 
a long-term time-averaged spectrum (all data) and a medium-term 
time-averaged spectrum (intensive monitoring only); total exposures were 
278.1 ksec and 100.4 ksec, respectively.

Response matrices were generated using {\sc pcarsp v.8.0} over several periods
between 2003 and 2005. The PCA response evolves slowly; due to the gradual 
leak of propane into the xenon layers, the spectral response hardens 
slowly with time. For example, application of a 2003 matrix to IC~4329a data 
observed in 2005 (or vice versa) results in a shift in photon index of 
only $\sim$0.04. For long-term time-averaged spectral analysis below, a 2004 
response matrix was used.

\subsubsection{HEXTE Data Reduction}    

The High-Energy X-Ray Timing Experiment (HEXTE) aboard \xte\ consists of
two independent clusters (A and B), each containing four scintillation 
counters (see Rothschild \et\ 1998) which share a common 1$\fdg$ FWHM 
field of view. Source and background spectra were extracted from each 
individual \xte\ visit using Science Event data and standard extraction 
procedures. The same good time intervals used for the PCA data (e.g., 
including Earth elevation and SAA passage screening) were applied to the 
HEXTE data. To measure real-time background measurements, the two HEXTE 
clusters each undergo two-sided rocking to offset positions, in this 
case, to 1.5$\arcdeg$ off-source, switching every 32 seconds. There is a 
galaxy cluster (Abell 3571, z = 0.039) located about 2$\arcdeg$ south 
of IC~4329a, at R.A.\ = 13h 47m 29s,  decl.\ = --32$\degr$ 52 m. This source 
is detected in the \xte\ all-sky slew survey (XSS; Revnivtsev \et\ 2004), 
which shows the 8--20 keV flux of this source to be about half that of 
IC 4329a.  This suggests that the presence of A3571 could affect HEXTE 
background determination for any HEXTE cluster which rocks to within 
roughly 1--2$\degr$ of the center of A3571 (accounting for the extended 
size of the galaxy cluster and the $\sim$1$\degr$ FWHM collimator 
response of the HEXTE clusters). However, 
%%%%%spectral modelling by Horner (2005) and De Grandi \& Molendi (2002) 
%%%% suggest A3571 has a 7--8 keV temperature, and 
{\it BeppoSAX}-PDS observations have shown no detection of A3571 above 
15 keV (Nevalainen \et\ 2004). The 8--20 keV emission seen by the XSS must 
therefore be emission only between 8 and $\lesssim$15 keV, and so the 
presence of A3571 can be safely ignored as far as contaminating the 
HEXTE background data is concerned. Cluster A data taken between 
2004 Dec 13 and 2005 Jan 14 were excluded, as the cluster did not rock 
on/off source. Detector 2 aboard Cluster B lost spectral capabilities
in 1996; these data were excluded from spectral analysis. Cluster A and B 
data were extracted separately and not combined, as their response matrices 
differ slightly. Deadtime corrections were applied, and the individual 
spectra were merged for each cluster to form a long- and medium-term 
time-averaged spectrum in the same manner as the PCA data. The net source 
exposure times for the long-term spectrum were 100.5 ksec (cluster A) 
and 101.8 ksec (cluster B). Net source exposure times for the medium-term 
spectrum were 37.0 ksec (A) and 36.5 ksec (B). 

\section{PCA/HEXTE Time-Average Spectral Analysis}

\setcounter{footnote}{0}

We first discuss spectral fits to the long- and medium-term \xte\ PCA
and HEXTE data. The energy resolution of the PCA is low, but we discuss 
these data first for the purpose of characterizing the broadband hard X-ray 
continuum and constraining the strength of the Compton reflection component, 
and applying those constraints to the higher resolution \xmm\ EPIC data.
In addition, model fits to the long- and medium-term time-averaged spectra 
were used as templates when analyzing the time-resolved spectra, as 
described in $\S$5. The medium-term data constitutes a large fraction 
(36$\%$) of the total long-term data, so the two spectra are not independent.
 
PCA data below 3.5 keV were discarded in order to disregard PCA calibration
uncertainties below this energy and to reduce the effects of the warm absorber
(e.g., S05). All PCA data above 24 keV were ignored:
there are background features between 24 and 32 keV
that are not properly modeled by the PCA 
calibration (as of 2006 Feb) 
for faint sources (C.\ Markwardt 2006, priv.\ comm.), and the 
source counts become dominated by statistical uncertainties at
energies higher than $\sim$38 keV in the long-term data
and $\sim$30 keV in the medium-term data. 
HEXTE data below 25 keV were excluded, as the responses of 
clusters A and B diverge somewhat for faint sources (N.\ Shaposhnikov 2005,
priv.\ comm.); there is good agreement above 25 keV. HEXTE data above 
100 keV were excluded as the source count rate becomes dominated by
statistical uncertainties at higher energies. For the long-term time-averaged 
spectrum, HEXTE data above 50 keV were grouped as follows: 
channels 51--60, channels 61--75, and channels 76 and above were binned 
in groups of 2, 3, and 4, respectively. For the medium-term data, data 
above 39 keV were also binned: channels 39--50 and
channels 51 and above were binned in groups of 3 and 8, respectively.

All spectral fitting for both \xte\ and \xmm\ data was done with 
XSPEC v.11.3.2 (Arnaud 1996). Errors quoted for spectral fit 
parameters are 90$\%$ confidence. All line energies quoted are for 
the rest frame unless otherwise indicated.
All spectral fits included a {\sc wabs} 
neutral absorption component. The Galactic column towards IC~4329a
is $4.42 \times 10^{20}$ cm$^{-2}$ (Dickey \& Lockman 1990),
but previous studies have suggested additional
absorption (e.g., Gondoin \et\ 2001). Our analysis excludes data
$<$2.5 keV (pn) or $<$3.5 keV (PCA) and is 
relatively insensitive to the total 
(Galactic + intrinsic) neutral column density; we
leave this parameter free in our fits.

%%%%%%update - see dothis_RXTE or dothis1225
%%%%    (not dothis_long2hexte)
                                      
We first discuss joint fits to the long-term PCA/HEXTE spectra, 
plotted in Figure 2. In these fits, a constant coefficient was included
to account for minor normalization offsets between the PCA and HEXTE.
The first model tested, a simple power-law modified by cold 
absorption  %%%PWC0823
(Model 1), yielded large residuals near 6 keV due to the presence of
the Fe K$\alpha$ line; $\chi^2$ was 2129.17/128 degrees of freedom 
($dof$). The addition of a Gaussian emission component (Model 2) at 
6.38 keV, the Fe K$\alpha$ line, was significant at greater than  %%%PGWC0823
99.99$\%$ confidence in an F-test, as $\chi^2$/$dof$ fell to
586.81/125. However, there still remained broad residuals due to
the presence of the Compton reflection component. Model 3 included a      %%%%  PPG0823
{\sc pexrav} component to model reflection of an underlying power-law 
component off optically thick and neutral material (Magdziarz \& Zdziarski 
1995). 
The inclination angle $i$ was assumed to be 30$\degr$;
the power-law cutoff energy was kept fixed at 270 keV as per the results 
of Perola \et\ (1999). Elemental abundances were kept fixed at solar.
Adding the {\sc pexrav} component proved significant at greater than 
99.99$\%$ confidence in an F-test, since $\chi^2$/$dof$ fell to 208.14/123.
The photon index $\Gamma$ was 1.894$^{+0.013}_{-0.015}$.       %%%%%  PPG1021
The reflection fraction $R$ 
(defined as $\Omega$/2$\pi$, where $\Omega$ is the solid
angle subtended by the reflector)
was 0.51 $\pm$0.04. (With HEXTE data 
omitted, $R$ from PCA data alone is 0.55 $\pm$ 0.07;
$\Gamma = 1.908 \pm 0.015$.) Due to the degeneracy between $R$ and $i$,
it was not possible to fit for both these parameters simultaneously, although we note
that for a more face-on inclination 
(cos($i$) = 0.95), $R$ decreased slightly to 
0.48 $\pm$ 0.04 (with $\chi^2$ increasing by only 0.1). 
With the constant coefficient of the PCA fixed at 1.0, 
the constants for the HEXTE A and B spectra were 0.90 and 0.92, respectively.
Table 1 summarizes the model parameters (including line emission parameters) 
for Model 3. Figure 2 shows the data/model residuals for the three models 
discussed.
                                                                           
Given the limit of PCA calibration (data/model residuals 
$\sim$2$\%$\footnote{See http://lheawww.gsfc.nasa.gov/users/keith/rossi2000/energy$\_$response.ps})
and the high signal-to-noise ratio of this spectrum, systematic errors
associated with background modeling and the PCA's spectral response dominated
the uncertainties. Finding a statistically acceptable time-averaged fit
(with reduced $\chi^2_{\nu} \sim 1$) was thus unlikely even if
the model was representative of the intrinsic spectrum of the source.
Additionally, given the low energy resolution of the PCA ($\sim$1~keV at 6~keV),
the PCA is largely insensitive to the detailed shape of the Fe K$\alpha$ line;
there were no other obvious residuals seen in the Fe K bandpass,
e.g., the Fe K$\beta$ emission line was detected only at 90$\%$ confidence,
and using a diskline model instead of a Gaussian yielded an equally acceptable fit.
Model 3 was therefore adopted as a satisfactory baseline 
fit to the long-term time-averaged PCA/HEXTE spectrum.
                                                          
Spectral fitting to the medium-term time averaged PCA/HEXTE spectrum     %%%% dothis_med_hex
proceeded in a manner identical to the long-term spectrum, testing 
the same three models, and with very similar results, although 
$\chi^2$/$dof$ was lower due to a lower signal-to-noise ratio. The results of 
the Model 3 fit are summarized in Table 1. Compared to the long-term 
time averaged spectrum, the medium-term $\Gamma$ and $R$ were both  
slightly lower. However, we note the similarity in line emission 
parameters. Furthermore, the fact that the same type of model was 
representative of both spectra is consistent with the notion that the 
general form of the broadband X-ray emission did not change over time.

Since the primary goal of the PCA/HEXTE fitting was to constrain the photon
index $\Gamma$ and the reflection strength $R$, we studed more 
detailed models only using the higher resolution 
EPIC-pn data ($\S$4). In particular, we adopted $R$ = 0.51, the reflection strength from
the long-term fits, to use in the EPIC-pn spectral analysis.

\section{pn Fits}     
%%%%  /n7/.../tr16/dothis1212,dothis1616_absn

We now discuss spectral fits to the time-averaged EPIC pn spectrum, 
shown in Figure 3 (all pn data are plotted with
rebinning every 15 bins unless otherwise specified). 
The best-fitting model was then used as a template 
for time-resolved spectroscopic studies in $\S$5. We ignored data $<$2.5 keV
and concentrated only on hard X-ray emission. 
$\S$4.1 describes preliminary fits to the data, including attempts to model
the Fe K emission using Gaussians. $\S$4.2 discusses 
relativistic diskline model fits. $\S$4.3 discusses model fits to
a narrow absorption feature seen at 7.68 keV, as well as
its detection significance and origin. In $\S$4.4, we present joint pn/{\it RXTE} fits.
In $\S$4.5, we show that the pn spectrum was
consistent with the MOS2 and simultaneous \xte\ spectra.

\subsection{Fe K$\alpha$ and K$\beta$ Emission: Preliminary Fits}

Data/model residuals to a simple power-law fit (Model 1)
are shown in Figure 3, the Fe K$\alpha$ core
and Compton reflection hump above 7 keV are clear.
Model 2 included a Gaussian near 6.4 keV.
For Model 3, a {\sc pexrav} component with $R$ fixed at 0.51 was added. 
A high-energy cutoff of 270 keV, solar metal abundances, and an inclination of 
30$\degr$ were assumed. 
Data/model residuals are shown in Figure 4, indicating an excess
near 7.0 keV, and a deficit near 7.7 keV suggesting
an absorption feature of some kind. 
The best-fit spectral parameters for Models 1--3 are listed in Table 2.
As an alternate parameterization of the Fe K edge, we fit 
an edge instead of a {\sc pexrav} component, finding an 
edge energy of 7.16 $\pm$ 0.07 and an optical depth of 0.05 $\pm$ 0.01, 
consistent with Gondoin \et\ (2001).

To model the emission near 7.0 keV (the ``blue peak''; 
``red peak'' refers to the 6.4 keV line) we added a second Gaussian (Model 4).
The best-fit had a blue peak with energy centroid at 6.93$^{+0.09}_{-0.10}$ keV 
and a width $\sigma_{7.0}$ of 121$^{+95}_{-59}$ eV.
It was significant at $>$99.99$\%$ in an F-test to add this component.
Data/model residuals are plotted in Figure 4;
the best-fit spectral parameters are listed in Table 2.

In this model, the red peak centroid energy, 6.39 $\pm$ 0.01 keV, was 
consistent with K$\alpha$ emission from neutral or mildly ionized 
Fe. The core width $\sigma_{6.4}$, 91 $\pm$ 13 eV, corresponds to a 
FWHM velocity of 9700 $\pm$ 1400 km s$^{-1}$;
this is consistent with the value from {\it Chandra-HETGS},
15100$^{+12400}_{-10000}$ km $s^{-1}$ (Yaqoob \& Padmanabhan 2004).
The 6.4 keV line was resolved at the pn resolution of 
$\sim$140 eV: fixing $\sigma_{6.4}$ to 1 eV resulted in large
data/model residuals while $\chi^2$ 
increased by 91. The blue peak was also resolved:
fixing $\sigma_{7.0}$ to 1 eV caused $\chi^2$ to increase by 6.5, 
significant at 99.1$\%$ confidence in an F-test.

Given the energy resolution of the pn, it was not obvious from these 
parameters whether the blue peak was due to neutral 
Fe K$\beta$ at 7.06 keV, \ion{Fe}{26} at 6.966 keV, 
or a blend of both. To test if it 
was consistent with being due solely to Fe K$\beta$, 
we fixed the blue peak Gaussian's centroid energy at 7.06 keV and fixed the  
the K$\beta$$/$K$\alpha$ flux ratio at 0.13. The width $\sigma_{7.0}$ 
was kept tied to $\sigma_{6.4}$. The best  
fit model (Model 5; see Table 2) was slightly worse than Model 4
($\chi^2$ increased by 7.57),
and, as shown in Figure 4, small data/model residuals appeared near 
6.8 keV, suggesting excess unmodeled emission on the red side of 
the blue peak.

The data/model residuals for Models 4 and 5 (see Figure 4) also
suggested unmodeled emission on the red side of the
K$\alpha$ line near 6.1 keV, prompting us to test for the presence of a 
Compton-scattering shoulder. Such a feature might be expected 
at 6.24 keV due to Fe K$\alpha$ photons Compton-scattering on 
electrons in the medium in which the line is formed.
The shoulder/core intensity ratio indicates
scattering in, and an origin for the Fe K$\alpha$ line in, 
Compton-thick or -thin material (ratio above or below 
$\sim$0.1--0.2, respectively; see Matt 2002 for further details).
Building on Model 4, we added a third Gaussian 
emission component with centroid energy fixed at 6.24 keV and
a width tied to that for the 6.4 keV core (Model 6). 
In the best-fit model, the equivalent width $EW$ was 9 $\pm$ 5 eV and the
centroid energy was 6.13$^{+0.07}_{-0.25}$ keV
(it was significant at 98.2$\%$ confidence in an F-test to
let the centroid energy not be fixed at 6.24 keV).
An F-test showed it was significant at 99.6$\%$ 
to add this component (compared to Model 4). 
Best-fit spectral parameters are listed in Table 3. 
Considering the pn energy resolution, this feature is 
consistent with emission from a Compton shoulder. 
The shoulder-to-core intensity in this case is 0.13$^{+0.10}_{-0.08}$, so
we cannot rule out either a Compton-thin or Compton-thick origin for the K$\alpha$ line.
In these models, the Fe K$\alpha$ line centroid energies are consistent with
emission from neutral or only mildly ionized Fe.

\subsection{Fe K$\alpha$ and K$\beta$ Emission: Diskline Fits}

We now discuss whether relativistic diskline models (Laor 1991)
can describe the Fe K profile of IC 4329a. 
Done \et\ (2000) noted from the {\it ASCA} observation
that IC 4329a seems to have a slightly broadened red wing, 
though not as broad as that of the ``archetypal'' broad-line source, 
MCG--6-30-15 (Tanaka \et\ 1995); see Figure 2 of Done \et\ (2000).
Their best-fit models suggested that 
there was not significant line emission at radii smaller than 
$\sim$30--100 $R_{\rm g}$ ($R_{\rm g}$ $\equiv$ G$M_{\rm BH}$/$c^2$).
However, compared to the {\it ASCA} data, the pn spectrum reveals 
additional emission and absorption features in the Fe K band; 
in this section we explore if diskline models can 
successfully account for all of these features. 

As seen by McKernan \& Yaqoob (2004), the heights above 
the power-law continuum of the red and blue peaks in the 60 ksec
HEG spectrum were approximately equal, prompting 
McKernan \& Yaqoob (2004) to test 
if the two peaks could be the horns of a single diskline. 
Their best-fit model was one with an inclination angle $i$ = 
$24^{+9}_{-1}$$\degr$, rest-frame line energy of 
6.74$^{+0.22}_{-0.13}$ keV, a flat radial emissivity profile
($\beta$ $<$ 0.7, where the radial emissivity per unit area is 
quantified as a power law, r$^{-\beta}$), and emission between 
$\sim$6 and 70 $R_{\rm g}$. However, the current EPIC-pn spectrum 
clearly shows that the $EW$ and height above the power-law continuum
of the red peak are much greater than those of the blue peak. 
A relativistic diskline model with near face-on inclination, 
e.g., $\lesssim$10$\degr$, can typically produce only 
one peak, and thus one diskline cannot account for both peaks 
simultaneously. A diskline model with intermediate or high 
inclination angles is required to produce clearly-resolved two peaks.
We attempted to fit such a model to the pn data,
using a {\sc Laor} diskline (Model 7).
We tested inclination angles of 30$\degr$, 50$\degr$, and 70$\degr$.
The outer radius $R_{\rm out}$ was fixed at 400 $R_{\rm g}$.
In all fits, the inner radius $R_{\rm in}$ 
went to very large values, usually 250--350 $R_{\rm g}$, 
with very poor constraints. All fits yielded poor or no constraints on 
$\beta$. In all cases, there were large systematic data/model 
residuals across the Fe K bandpass, as
shown in Figure 4 (see also Table 4). The values of $\chi^2$
for best-fit models were typically 1700--1920 for 1691 $dof$, 
significantly worse fits than e.g., the double-Gaussian model.  
We conclude that a single-diskline model is an inaccurate 
description of the data.

The unmodeled excess emission near 6.1 keV and 6.8 keV in the 
double-Gaussian model (K$\beta$ energy fixed; Model 5) 
potentially signified that the 
K$\alpha$ and K$\beta$ emission components could each be independently 
modeled better with a diskline instead of symmetric Gaussian. 
McKernan \& Yaqoob (2004) tested dual-diskline models for the 
HEG spectrum; the best-fit model was one with a nearly face-on disk,
$i$ $<$ 6$\degr$, a moderately steep emissivity index 
$\beta$ = 2.4 $\pm$ 0.2, and emission spanning the radii between 
6 $R_{\rm g}$ (fixed) to 600 $R_{\rm g}$. S05's best-fit 
double-diskline model was found to be consistent with this, with $\beta$ = 1.4, $i$ $<$ 17$\degr$,
and a profile very similar to a double-Gaussian model.
We also fit a double-diskline model to the data, with $i$, 
$\beta$, $R_{\rm in}$, and $R_{\rm out}$ for the red and blue 
disklines set equal to each other, and the normalization 
of the blue diskline equal to 0.13 that of the red diskline. 
$R_{\rm out}$ was kept fixed at 400 $R_{\rm g}$; it was not significant
to thaw this parameter. Our best-fit model (Model 8, see Table 5) agrees 
generally well with that of S05; we found $i$ $<$ 12$\degr$,
$\beta$ = 2.1 $\pm$ 0.3, and $R_{\rm in}$ $<$ 26 $R_{\rm g}$.
The energy of the red peak was 6.44$^{+0.02}_{-0.01}$ keV, more 
consistent with K$\alpha$ emission from mildly ionized Fe than from 
neutral Fe. The blue peak energy was 6.98 $\pm$ 0.09 keV.
The equivalent width $EW_{\rm K\alpha}$ was 86$^{+17}_{-8}$ eV.
The values of $EW_{\rm K\alpha}$ and Compton reflection strength $R$
were consistent with the prediction of George \& Fabian (1991) that, for 
reflection off neutral material, the reflection strength $R$ will be equal to 
$EW_{\rm K\alpha}$ / 150 eV. 

For this model, $\chi^2$/$dof$ was 1612.65/1688, 
similar to that for the Compton shoulder model.
As shown in Figure 4, the dual-diskline model accounts well for all of the
excess emission-like residuals near 6.1 and 6.8 keV, 
suggesting that the dual-diskline model is a better description 
of the data than the double Gaussian models (without a Compton shoulder). 
To quantify this 
statement, we replaced only the red peak Gaussian in Model 5 
with a {\sc Laor} component and refitted; $\chi^2$/$dof$ fell to 
1621.18/1688, significant at 97.3$\%$ confidence in an F-test. 
Replacing only the blue peak Gaussian with a {\sc Laor}
component and refitting yielded $\chi^2$/$dof$ = 1623.74/1687, significant at
only 84.8$\%$ confidence in an F-test.

A {\sc Laor} profile assumes a maximally rotating Kerr black hole.
Substituting the {\sc Laor} model components with {\sc diskline} 
model components and assuming a nonrotating Schwarzschild 
black hole (Fabian \et\ 1989) yielded a virtually
identical fit to the dual-{\sc Laor} model
For the remainder of this paper, ``diskline''
will denote the {\sc Laor} model only.

%%%%% Another possible diskline model is one wherein the Doppler horns are very 
%%%%% close together and unresolved by the pn. A flat emissivity profile 
%%%%% and an inclination angle of $\sim$20--30$\degr$ can yield such a 
%%%%% profile for certain ranges of $\beta$ and $R_{\rm in}$. Additionally, 
%%%%% the blue horn in this profile is slightly higher than the red horn; 
%%%%% the fact that the best-fitting rest-frame energy for the Fe K$\alpha$ diskline
%%%%% in Model 8 was 6.44 $\pm$ 0.01 suggests that this model is worth exploring.
%%%%% Inclination angles much higher than $\sim$30$\degr$ yield line profiles 
%%%%% with Doppler peaks that could be resolved by the pn. We tried applying 
%%%%% such a model to the data, fixing the inclination angle to values in the
%%%%% range 20--30 $\degr$. However, these models could not account for the peak of 
%%%%% the emission and the broad red wing simultaneously. Values of 
%%%%% $\chi^2$/$dof$ generally stayed about 1900/1891, and we conclude that
%%%%% this model does not describe the emission profile as well as Model 8 above.

Finally, we explored the possibility that in addition to the
diskline emission, there could be non-negligible emission
from narrow components. We added two narrow Gaussians to Model 8 at the 
rest-frame energies for Fe K$\alpha$ and K$\beta$, with the 
K$\beta$ normalization set to 0.13 times that of the K$\alpha$ line. 
We first considered the possibility that each broad component
is described by a low-inclination diskline component. 
We kept the widths of both Gaussians fixed at 10 eV.
The best-fit model (referred to as Model 9 hereafter) 
was one with diskline parameters nearly identical
to those in Model 8, with the K$\alpha$ and K$\beta$ lines
contributing only a small amount, and formally consistent with
upper limits only in $EW$. The best-fit parameters are listed in Table 6.
Data/model residuals appeared identical to those in Model 8.
A key point here was to determine
the maximum allowable $EW$ for the narrow K$\alpha$ line without 
significantly changing any of the diskline parameters; this $EW$ was 30 eV,
and in this case, the $EW$ of the broad K$\alpha$ line was 51 eV.
We conclude that if the dual broad components are described by 
low-inclination, moderately-relativistic 
disklines, then the bulk or entirety of the emission
must be from the broad components, but a contribution
from dual narrow components cannot be ruled out.

We next considered a model wherein high-inclination disklines are responsible 
primarily for the red-wing emission, with a narrow component 
accounting for the bulk of the peak emission very close to 6.400 or 7.06 keV.
We kept the rest-frame energies of the Fe $\beta$ broad and narrow 
components fixed at 7.06 keV. Parameters for the best-fit model 
(henceforth referred to as Model 10) are listed in Table 6. The 
best-fitting $\chi^2$ was nearly identical to that for Model 8 (no narrow
emission lines). Data/model residuals looked identical to those in Model 8. 
The width $\sigma$ of the narrow K$\alpha$ line, 
66$^{+19}_{-22}$ eV, corresponds to a FWHM velocity of 7100$^{+2000}_{-2400}$ km s$^{-1}$. 
For Keplerian motion, this corresponds to a radius of 
$\sim$900 $R_{\rm Sch}$ (1 $R_{\rm Sch}$ $\equiv$ 2$G$$M_{\rm BH}$/$c^2$). 

We conclude that if a relativistic diskline is indeed required 
to best model the emission, then a degeneracy is present:
models containing two broad lines and two narrow lines
can fit the observed profile of IC 4329a equally well using
either a low-inclination diskline (plus little or no narrow emission)
or by narrow components plus high-inclination disklines.
Alternatively, the Fe K emission profile can be described
just as well using Gaussians for the 6.4 and 7.0 keV peaks
plus a Compton shoulder with $EW$ $\sim$10 eV.
Below, we adopt Model 8, the dual-diskline model with
no narrow emission lines, as our 'baseline' model of the
emission profile shape.

To further quantify any possible narrow emission or absorption features  
in addition to the Fe K$\alpha$ and K$\beta$ profiles, we 
added a narrow Gaussian to Model 8, and, using the {\sc steppar} 
command in XSPEC, derived confidence contours of line intensity 
versus centroid energy. The results are shown in Figure 5, and demonstrate that
there were no additional obvious emission or absorption signatures from
Fe {\sc XXV} or {\sc XXVI} at their rest-frame energies. 
In fact, the only obvious feature was the narrow absorption feature near 7.7 keV.
Fitting narrow Gaussians at the %%%%% \ion{Fe}{25} or \ion{Fe}{26} 
rest-frame energies yielded upper limits in 
$EW$ to \ion{Fe}{25} emission (2 eV),
\ion{Fe}{25} absorption (3 eV),
\ion{Fe}{26} emission (15 eV),
and \ion{Fe}{26} absorption (8 eV).
The constraints on line emission are stronger
than those found by Bianchi \et\ (2005) using the 
10 ksec \xmm\ observation of IC 4329a in 2001.

\subsection{The 7.68 keV Narrow Absorption Feature}

We now focus on modeling the narrow absorption feature at
$\sim$7.7 keV, which persisted in the data/model 
residuals independently of how the emission lines were modeled, suggesting
a physical origin distinct from that of the 
emission lines. $\S$4.3.1 discusses simple model fits to the absorption line.
In $\S$4.3.2, we show that this feature is likely not an instrumental artifact, 
and in $\S$4.3.3 we discuss the statistical 
significance of detecting this feature. $\S$4.3.4 discusses 
possible origins for the line.

\subsubsection{Gaussian Model Fit for the Line}

We started with Model 8, the dual-diskline model (though  
results were independent of which model was used to model the emission profile). 
We first tested the hypothesis that the feature was indicative of   %% PEXRIV1215.xcm
reflection off ionized material, and could be modeled by a K-shell edge
due to moderately ionized Fe. Adding an edge (Model 11) 
yielded a best-fit edge energy of 7.29$^{+0.25}_{-0.17}$ keV 
and an optical depth $\tau$ of 0.03 $\pm$ 0.01 (see Table 7). 
However, the data/model 
residuals near 7.6 keV were not strongly affected. Forcing the 
edge energy closer to 7.4--7.6 keV did not improve the
data/model residuals. 
Models incorporating a {\sc pexriv} component  
(i.e., in addition to the {\sc pexrav} component)
with $\xi$$\sim$3--10 erg cm s$^{-1}$
($\xi$ $\equiv$ 4$\pi$$F_{\rm ion}$/$n$;
$F_{\rm ion}$ is the 0.5--20 keV ionizing continuum flux;
$n$ is the density of the reflecting material) similarly proved 
unsuccessful.
We conclude that the feature is too narrow to be an 
Fe K absorption edge. The lack of an obvious Fe L edge due to
moderately ionized Fe also argues against the edge hypothesis.

We then attempted to model the feature using a simple narrow, inverted 
Gaussian. We returned to Model 8, and added an inverted Gaussian near 
7.7 keV (Model 12). The best-fit model successfully removed the data/model 
residuals, as shown in Figure 4. Spectral parameters for the best-fit 
Model 12 are listed in Table 7. The best-fit rest-frame energy centroid 
was 7.68$^{+0.04}_{-0.03}$ keV (7.56$^{+0.04}_{-0.03}$ 
keV in the observed frame); $\sigma$ was $<$100 eV. 
The absolute value of the line intensity $\vert$$I$$\vert$ was 
$9.6 \pm 3.4 \times 10^{-6}$ ph cm$^{-2}$ s$^{-1}$; 
$\vert$$EW$$\vert$ was 13 $\pm$ 5 eV. The value of $\chi^2$/$dof$ was 
1589.46/1685. Compared to Model 8, 
$\chi^2$ dropped by 23.19, and an F-test for adding this component 
yielded a null hypothesis probability of $2.0 \times 10^{-5}$. 
However, there was no {\it a priori} reason
to expect an absorption feature at that energy, and so the 
F-test, when used in this ``standard'' manner, tends to overestimate 
the significance of detecting such a feature. The statistical significance 
of detecting this component will be addressed in more detail in $\S$4.3.3.

%%%% Finally, having properly quantified the emission and absorption
%%%% profile, we investigated the Fe abundance by studying the height of the 
%%%% 7.1 keV edge relative to the Compton reflection strength. In the 
%%%% {\sc pexrav} model, the Fe abundance had been kept frozen at 1.0, 
%%%% using the Anders \& Grevesse (1989) tables. Keeping $R$ frozen at 
%%%% 0.51 and thawing the Fe abundance made no change 
%%%% in the fit statistic; the best-fit Fe abundance was 1.0$^{+1.0}_{-0.4}$.
%%%% Using the Grevesse \& Sauval (1998) tables, the best-fit Fe 
%%%% abundance was 1.5 $\pm$ 0.7.

\subsubsection{PN Background/Instrumental Effects}    

The pn background spectrum shows no obvious features at 7.56 keV in the
observed frame. However, two instrumental features, a Ni K$\alpha$ 
emission line at 7.48 keV and a Cu K$\alpha$ emission line at 
8.05 keV, are expected (Katayama \et\ 2004; Freyberg \et\ 2004).
These lines originate from the electronics board located below the CCD camera.  
One might initially suspect that the 7.56 keV (observed-frame) 
absorption feature could be an artifact resulting from the 
7.48 keV line, but there are several arguments 
against this notion.  First, the background count rate is a factor 
of $\sim$40 fainter than that of the source:  
5--9 keV count rates in the background and source (after 
background subtraction) were 0.045 c s$^{-1}$ and 1.748 c s$^{-1}$, 
respectively. As shown in Figure 6, there are no obvious features
at these energies in the pn background spectrum of IC 4329a;
S05 attempted to detect 
expected instrumental lines in the pn background to verify the
energy scale, but did not detect any with sufficient significance.
Second, there were no differences seen in the 
source spectrum when screened against the periods of highest background rates, 
as discussed in $\S$2. Third, if it were the case that the 7.48 keV 
line affects the source spectrum, then one might also expect 
a feature near 8.05 keV as well, since the 8.05 keV line is actually 
stronger (Katayama \et\ 2004). However, no such features are evident 
in the source spectrum: adding narrow Gaussians in either emission or
absorption at 7.48 or 8.05 keV had no effect on the fit or residuals. 
Finally, these instrumental lines' fluxes are 
at a minimum within 5$\arcmin$ of the core  (Katayama \et\ 2004); the 
background in this case was extracted over a region located 3$\arcmin$ 
away from the core.
%%%%%%When the pn is in small window mode, the exposed part of the CCD is above the venting hole and contamination is minimized.
We are confident in concluding that the feature 
at 7.56 keV in the observed frame is intrinsic to the spectrum 
of IC~4329a and is not an artifact of the instrument or the background.

It has been reported that for pn ``double event'' data, events
where photon energy is deposited on adjacent
pixels, the registered photon energy can be 20 eV greater than the energy
corresponding to an otherwise-identical single-pixel event 
(e.g., Pounds \& Page 2005). The pn data had initially been 
extracted using pattern 0--4 events (single and double events);
to test for this effect, we re-analyzed the pn data using only 
single-event (pattern=0) data. For both filtered and unfiltered 
data, the pattern=0 spectrum was virtually identical to the 
pattern$\leq$4 spectrum, though with $\sim$30$\%$ fewer counts, 
and a continuum slope that was steeper by $\sim$0.05 in photon index. 
Importantly, however, the fitting results for Fe K bandpass features 
were virtually identical in all cases. We conclude that the emission 
and absorption features modeled here are not artifacts of pattern 
selection.

\subsubsection{Estimating the Significance of the Detection of the 7.68 keV Absorption Feature}

The standard, two-parameter F-test for the addition of a 
Gaussian to model the absorption line at 7.68 keV yielded 
a null hypothesis probability 
of $2.0 \times 10^{-5}$ (4.3$\sigma$ significance).
However, using the F-test in this manner
has a tendency to overestimate the detection significance, as the F-test
does not take into account the possible range of energies where a line might 
be expected to occur, nor does it take into account the number of bins
(resolution elements) present over that energy range. 
The F-test can yield the probability $P_1$ (equal to one minus the
null hypothesis probability) of finding a feature at
a given energy {\em if the line energy is known in advance};
see e.g., warnings by Protassov \et\ (2002).
However, in this case, there was no {\it a priori} expectation of a 7.68 keV
feature, and in searching for narrow features at arbitrary energies, one is
searching over many resolution elements and needs to account for
the possibility that narrow features can occur by chance due to 
statistical noise. One might search the 4--9 keV bandpass
for Fe K emission/absorption lines, including narrow features 
near 5--6 keV that have been generally interpreted as gravitationally 
redshifted Fe K emission (e.g., Turner \et\ 2002, 2004).
There are about 36 pn resolution elements over this energy range.
The probability of detecting a feature at any energy in this 
range due is found from $(P_1)^N$, where $N$ is the number of resolution 
elements. For IC~4329a, the probability that the
7.68 keV feature is spurious thus becomes $7.2 \times 10^{-4}$
(3.4$\sigma$ significance).
We note that if the spectral bin sizes are smaller than the instrument 
resolution, then the number of bins where a line may be located has probably 
been over-estimated; this means the detection probability may have 
been underestimated.

Monte Carlo simulations were performed as an additional test
of the line significance (see Porquet \et\ 2004).
These simulations tested the null hypothesis that the
spectrum is well-fitted by a model that does not include the
7.68 keV absorption feature.
For simplicity, Model 8 (dual disklines) was initially assumed. 

We generated 1000 fake spectra as follows: We ran {\sc fakeit none} on the null hypothesis model,
with the photon statistics for a 93.2 ksec exposure, and then re-fit the model to this new fake
spectrum, yielding a "modified" null hypothesis model. We ran {\sc fakeit none} a second time, 
using this re-fit model, again with a 93.2 ksec exposure. This was done to account
for the uncertainty in the null hypothesis model itself; such uncertainty is relevant, for example,
when one is testing for the presence of broad features,
when broad features are present in the null hypothesis model,
or when the original data set is relatively noisy.
This entire process was repeated to generate 1000 
fake spectra. 
For each fake spectrum, we re-grouped to a minimum of 50 counts bin$^{-1}$ and
re-fit the null hypothesis model,
obtaining a $\chi^2$ value (hereafter
$\chi^2_{\rm null}$). We then added a narrow Gaussian component
($\sigma$ = 10 eV) to the fit. The line normalization
was allowed to be positive or negative, to account for the possibility 
of detecting either an emission or absorption feature, since there had 
been no {\it a priori} expectation for either. 
We searched over the energy ranges 4.0--6.1 keV and
7.1--10.0 keV.  The range 6.1--7.1 keV was excluded
to avoid the possibility that the disklines or
7.1 keV {\sc pexrav} edge would bias the Monte Carlo results.
For each fake spectrum, we stepped the Gaussian centroid energy 
over this range in increments of 0.1 keV (e.g., slightly higher than the instrument resolution), 
fitting separately each time to ensure that the lowest  
value of $\chi^2$ was found. For each fake spectrum,
the minimum value of $\chi^2$ was compared with $\chi^2_{\rm null}$, yielding
a simulated $\vert$$\Delta\chi^2$$\vert$ value. Repeating this process 1000 times
yielded a cumulative frequency distribution of the $\vert$$\Delta\chi^2$$\vert$
expected for a blind line search, assuming the null hypothesis model is correct.
We compared this distribution with the observed $\vert$$\Delta\chi^2$$\vert$ value, 23.19 in
this case. Not a single fake spectrum had
$\vert$$\Delta\chi^2$$\vert$ $\geq$ 23.19.
The inferred probability 
that the null hypothesis model was correct and that the feature is due to
photon noise is thus $<$0.1$\%$; i.e., the 7.68 keV absorption feature is 
detected at $>$99.9$\%$ confidence.

%%%%We additionally searched over the energy ranges
%%%%2.5--4.0 keV, and 10.0--12 keV. We would generally not expect to see
%%%%strong emission from Fe or any other metal, 
%%%%with the exception of H-like S emission
%%%%at 2.62 keV (rest frame; 2.58 keV observed frame). However, we explored the
%%%%probability of detecting spurious narrow features due to statistical noise.
%%%%Monte Carlo simulations showed that $\vert$$\Delta\chi^2$$\vert$ of 18.9 was exceeded
%%%%XXX times. Therefore, if one considers these energy ranges to be as
%%%%relevant as the 4--9 keV range, then the significance of
%%%%the 7.68 keV feature drops to XXXX.

Similarly, these simulations also can be used to 
quantify the probability that the Compton shoulder
could be due to photon noise. The decrease in $\chi^2$ when the Compton shoulder
was added was 19.1 (comparing models 5 and 6). 
Monte Carlo simulations, run in the same manner as above, using Model 5 
(dual-Gaussian, Fe K$\beta$ energy fixed) as the
'baseline' model, and searching over 5.8--6.3 keV, yielded
no value of observed $\vert$$\Delta\chi^2$$\vert$ exceeded in 1000 simulations.
The probability that the Compton shoulder 
is due to photon noise is thus $<$0.1$\%$.

Finally, we briefly address ``publication bias'': there are $\gtrsim$
half a dozen \xmm\ observations of Seyfert 1 galaxies of length and quality
(in terms of total number of photons) similar to that of the current observation 
(e.g., MCG--6-30-15, Fabian \et\ 2002; NGC~3783, Reeves \et\ 2004). 
One could thus argue that the significance of 
detecting the 7.68 keV feature should be
reduced to $\gtrsim$99.4$\%$.  We leave it to the reader
to include the relevance of such publication bias.
However, given the very high value of $\vert$$\Delta$$\chi^2$$\vert$,
the unlikeliness that the feature is an artifact of background modeling, and
and the low probability that it could be due to photon noise, 
we henceforth assume that the 7.68 keV absorption line is intrinsic to IC 4329a
and adopt Model 12 as our new baseline model.

\subsubsection{The Origin of the Line}   

As far as the origin of this feature is concerned, there are several 
candidates to consider. K-shell absorption due to \ion{Fe}{25} 
or \ion{Fe}{26} is a prime candidate, given that absorption lines 
attributed to highly ionized Fe have been found in other AGN.
If the 7.68 keV line is due to \ion{Fe}{26}, then its
blueshift suggests an origin in material that is flowing towards
the observer with a velocity
$\sim$31000 km s$^{-1}$ relative to systemic.
If the absorption is due instead to the \ion{Fe}{25} resonance
line (6.697 keV in the rest frame), then the velocity
is $\sim$44000 km s$^{-1}$ relative to systemic.

To shed additional light on the origin of this feature, 
we modified Model 8 (dual-diskline)
by including an {\sc xstar} component (Bautista \& Kallman 2001)
to model absorption due to ionized material along the line of sight (Model 13).
We assumed solar abundances, and kept the redshift as a free
parameter. The best-fit model (see Table 7) was one with an ionization
parameter $\xi$ characterized by 
log($\xi$) = 3.73$^{+0.15}_{-0.13}$ erg cm s$^{-1}$.
At this ionization level, Fe K absorption is due almost exclusively to
\ion{Fe}{26}. 
The value of $\chi^2$/dof was 1584.92/1685, a slightly better fit compared
to Model 12. The column density 
required was $1.4^{+1.0}_{-0.5} \times 10^{22}$ 
cm$^{-2}$, and the radial velocity shift relative to systemic was
--0.093$^{+0.006}_{-0.002}$$c$, or $\sim$--27000$^{+2000}_{-1000}$ km s$^{-1}$.
The best-fit model is thus consistent
with the interpretation above that the absorption feature has an origin in
highly ionized material outflowing at $\sim$0.1$c$ relative to systemic.

It is not unreasonable to expect supersolar metallicities
in AGN environments, given that studies of the BLRs 
of quasars spanning a range of redshifts 
generally yield metallicities near or several times solar
(e.g., Dietrich \et\ 2003a, 2003b).
Our fit used solar abundances, but was 
rather insensitive to the Fe abundance, as 
one unresolved absorption line cannot break the
$N_{\rm H}$--abundance degeneracy. For instance, refitting with the
Fe abundance fixed at 3$\times$ solar yielded
$N_{\rm H} = 1.1^{+0.6}_{-1.0} \times 10^{22}$ cm$^{-2}$
(no change in $\chi^2$).

It is unlikely that the 7.68 keV feature is due to
K$\beta$ absorption by highly ionized Fe (greater than
\ion{Fe}{17}), otherwise one would expect strong K$\alpha$
absorption by the same species. 
For example, the energy of 7.68 keV 
is close to the absorption energy for
Fe K$\beta$ {\sc XXIII} or {\sc XXIV} at the systemic velocity,
with an ionization parameter log$\xi$ $\sim$ 2.5--3.
In that case, however, we would expect an absorption trough
due to all the K$\beta$ transitions from Fe {\sc XVII--XXIV},
between roughly 7.3 and 7.9 keV, and, moreover, 
a very strong K$\alpha$ absorption trough near 6.5--6.6 keV
which is not seen in the data. 
We can similarly rule out redshifted 
Fe K$\beta$ {\sc XXV} or {\sc XXVI} resonant
absorption (redshifts of 0.024 and 0.075 relative
to systemic, respectively). In these cases, we would expect to see strong
Fe K$\alpha$ {\sc XXV} or {\sc XXVI} resonant
absorption at similar redshifts. However, there are
no obvious data/model residuals 
at those energies, and forcing a narrow Gaussian at those
energies yields small upper limits on absorption,
$\vert$$EW$$\vert$ $<$ a few eV for either case.

It is unlikely that the 7.68 keV feature is due to
blueshifted K$\beta$ absorption from lowly ionized Fe ($\sim$7.1--7.2 keV
in the rest frame). This would require
the absorbing gas to have an ionization
parameter smaller than log$\xi$ $\sim$ 2.1.
%%%       (blueshifts of 0.065 - 0.080, or 19,000-24,000 km/sec)
For an absorber with log$\xi$ = 2.0 and a blueshift of 0.068$c$
relative to systemic,  a column density
of $1.2 \pm 0.2 \times 10^{22}$ cm$^{-2}$ 
is required to fit the 7.68 keV line.
However, this model introduces very strong spectral curvature
below $\sim$4 keV; such curvature is not seen in the EPIC spectrum, as
{\sc XSTAR} modeling using low ionization photo-ionized absorbers
yield poor fits to the pn spectrum below $\sim$4 keV for 
column densities above 
roughly 10$^{21}$ cm$^{-2}$ (as shown in Figure 7a).  
Additionally, for 
log$\xi$ in the range $\sim$ 1.5--2.1, Fe K$\alpha$ absorption is just as 
strong, or stronger than, Fe K$\beta$ absorption.
If the 7.68 keV line were from lowly ionized Fe K$\beta$,
then we would expect Fe K$\alpha$ absorption at the same blueshift, near
6.9 keV (source frame). There are no obvious residuals near this energy; 
forcing an inverted narrow Gaussian into the model 
yields no change in the fit, with $\vert$$EW$$\vert$ $<$ 2.8 eV.
One can alleviate the requirement of having
Fe K$\alpha$ absorption by noting that
K$\alpha$ absorption is negligible compared to 
K$\beta$ for log$\xi$ $\lesssim$ 1.2. In that case,
the dominant species will be $<${\sc XVII}; the L-shell is full,
so K$\alpha$ absorption cannot be produced. However, as above, 
the required column density would introduce 
very strong spectral curvature into the spectrum below 
$\sim$4 keV, which is not seen. An origin for the 7.68 keV    
absorption feature in Fe K$\beta$ thus seems unlikely.

We next considered an origin in
very highly blueshifted, moderately ionized Fe K$\alpha$
(Fe $\sim$ {\sc XVII--XXII}, rest-frame energies $\sim$6.44--6.51 keV). 
For instance, an {\sc xstar} absorber with 
log$\xi$ = 2.1 $\pm$ 0.2, a column density 
$N_{\rm H} = 1.8 \pm 0.5 \times 10^{20}$ cm$^{-2}$,
blueshifted by 0.16 $\pm$ 0.01 $c$ relative to systemic provided almost 
as good a fit above 3 keV as Model 13,
with $\chi^2$/$dof$ = 1597.41/1685.  
Absorption due to a range of Fe species creates a trough;
for this low a column, the trough is not deep, but
at the resolution of the pn, the absorption feature
has $EW$ $\sim$ 12 eV and yielded a similar fit to a narrow
\ion{Fe}{26} line. However, extrapolation of this model to 
energies down to $\sim$0.9 keV revealed large data/model discrepancies. 
Specifically, one expects very strong 
($EW$ = 25--30 eV) absorption features near rest-frame energies of 
$\sim$0.91--0.94 keV due to L-shell absorption by moderately ionized 
Fe ({\sc $\sim$ XIX--XX}) and near 1.86 keV, likely due to \ion{Si}{14}. 
For an absorber blueshifted by 0.16$c$, we expect
such features to appear at 1.06 and 2.13 keV, but the pn data 
at those energies show no such features ($\vert$$EW$$\vert$ $<$ 1 eV). More generally,
the lack of {\it any} strong, highly blueshifted Fe L absorption lines and edges (S05) 
argues against intermediate species of Fe.
An origin in very highly blueshifted \ion{Fe}{17} thus seems unlikely.

There are transitions around
7.6--7.8 keV associated with highly ionized Ni K
({\sc XXV -- XXVI} or so), but an origin in Ni K with a velocity
near systemic is unlikely
unless Ni is many times overabundant compared to Fe in IC~4329a.
%%%%
%%%%we would expect to see very strong evidence of additional 
%%%%absorption due to highly ionized
%%%%Fe K$\alpha$ in the $\sim$6.6--6.96 keV range.
%%%%Even if the 7.68 keV feature is due to strongly 
%%%%redshifted He- or He-like Ni or strongly blueshifted
%%%%low- or intermediate-ionized Ni, we would still expect 
%%%%strong evidence for additional Fe K absorption lines 
%%%%at the appropriate redshift or blueshift. 
%%%%Given that there is only one absorption line strongly
%%%%evident in the data, an origin in Ni seems unlikely.

A local (z=0) origin should be discussed as well, given that
several recent soft X-ray AGN spectra have revealed evidence 
for absorption due to the local hot interstellar medium 
(e.g., McKernan, Yaqoob \& Reynolds 2004).
In the case of IC~4329a, a local origin for the line would require
an origin in K$\beta$ absorption from 
intermediate ionized Fe ($\sim$ {\sc XXII}).
However, an accompanying K$\alpha$ line would be expected in this case.
Furthermore, the very high column density required to produce such a line
($\gtrsim$10$^{22}$ cm$^{-2}$) is an implausible physical description of the
local hot interstellar medium.  
Similarly, the relativistic velocities
required for strongly blueshifted Fe K$\alpha$ absorption
are not associated with the local hot interstellar medium. We
conclude that a local origin for the line is unlikely.

Finally, Figure 7b shows a model spectrum of the effect of 
the soft X-ray absorbing components seen in S05 on a simple power-law
continuum. These low- and moderately ionized components are too low in column
density to introduce strong spectral curvature above $\sim$4 keV,
suggesting that our emission and absorption modeling of the Fe K 
bandpass is robust.

%%%%% KS sez: No Kbeta    You can fairly safely rule out the 
%%%%% Fe Kb interpretation, from the already
%%%%% pretty deep low ionization/neutral absorption seen in the soft X-ray
%%%%% band, you would not expect to see even a trace of the Fe Kb line.

\subsection{Joint PCA/HEXTE/pn Fits}

We applied our best-fitting {\sc XSTAR} model to 
the pn and long-term {\it RXTE} spectra, fitting simultaneously.
Constant offsets were included to account for cross-instrument normalizations.
When the photon indices for all three instruments were tied,
we obtained a best-fit with $\chi^2$/$dof$ = 2140.67/1808    %%%%JOIN004
and $\Gamma$ = 1.827 $\pm$ 0.008.
Allowing the photon index for the pn to differ
resulted a much better fit, as $\chi^2$/$dof$ dropped to $\sim$1848/1807.
The best-fit model had a {\sc Laor} inclination of 10 $\pm$ 7$\degr$; 
we then re-fit with the {\sc pexrav} inclination fixed at 10$\degr$
The best-fitting spectral values are listed in Table 8.

\subsection{Consistency Checks} 

We checked for consistency between the pn and the MOS2
by applying the above models to the MOS2 data between 2.5 and 12 keV.
Figure 8 shows the pn and MOS2 residuals to models consisting of
a power-law plus {\sc pexrav}, modified by neutral absorption;
the shape of the Fe K profiles are overall similar.
Applying the models above to the MOS2 data yielded
quite similar fits, with
most spectral parameters consistent between the two instruments,
though the photon indices for the MOS2 spectra were about 0.15 steeper.
We were able to confirm that the Fe K$\beta$ line is detected in the MOS2 data
at $\sim$90$\%$ confidence, using an F-test to compare models 3 and 4. 
However, due to the lower signal-to-noise ratio in the MOS2, it was not possible to
unambiguously verify the existence of the 7.68 keV 
absorption line ($\vert$$EW$$\vert$ $<$ 29 eV)
or confirm that a dual-diskline model was a better fit than
a double-Gaussian model.

To check for consistency between the pn and the \xte\ PCA, we 
generated a ``7 day'' simultaneous
PCA spectrum using \xte\ data taken from the 36 obsids that occurred
within $\pm$3.5 days of the midpoint of the \xmm\ observation.
This resulted in a PCA spectrum with a good exposure time of 19.6 ksec,
enough to 
confirm that an Fe K$\alpha$ line is present, but not enough to detect
other Fe K bandpass spectral features. The photon index
for Model 3 (Fe K$\alpha$ line only), applied to the ``7 day'' spectrum, was
1.89 $\pm$ 0.04, steeper than that of the model when applied to the pn data.
The 2--10 keV flux from the
``7 day,'' medium-term and long-term modeled spectra are
$11.0 \times 10^{-11}$ erg cm$^{-2}$ s$^{-1}$,
$11.2 \times 10^{-11}$ erg cm$^{-2}$ s$^{-1}$,
and $12.0 \times 10^{-11}$ erg cm$^{-2}$ s$^{-1}$, respectively.    
These values are about 30$\%$ higher
than the 2--10 keV flux estimated from model fits to the \xmm\ pn
spectrum, $8.9 \times 10^{-11}$ erg cm$^{-2}$ s$^{-1}$.
This discrepancy with the PCA normalization
has been noted previously; see, e.g., 
Yaqoob \et\ (2003)\footnote{This issue is also discussed in an {\it ASCA} Guest Observer Facility
Calibration memo at \\
http://heasarc.gsfc.nasa.gov/docs/asca/calibration/3c273\_results.html}.

\section{Time-Resolved Spectral Fitting}

Studying the time-resolved variability of the Fe K line
and continuum components provides a complementary analysis to single-epoch
spectral fitting. The simplest models predict that if
the reprocessing time scale is negligible, then variations
in the Fe K line flux should track continuum variations,
modified only by a time delay equal to the light travel time between
the X-ray continuum source and the Fe K line origin.
However, recent studies of both broad and narrow Fe K lines
have not been able to support this picture (e.g.,
Iwasawa \et\ 1996, 1999; Reynolds 2000;
Vaughan \& Edelson 2001, Markowitz, Edelson \& Vaughan 2003).
Most of these studies found that Fe K
lines tend to vary much less than the continuum,
with no evidence for correlated variability, on any time scale.
Markowitz, Edelson \& Vaughan (2003) additionally
showed that Fe K lines, like the continuum, tend to exhibit stronger
variability amplitudes towards relatively longer time scales.

Variability of narrow absorption features has also been reported, 
e.g., Reeves \et\ (2004) found evidence for the narrow \ion{Fe}{25} 
absorption line
in the {\it XMM-Newton} spectrum of NGC 3783 to increase in equivalent
width over $\sim$a day, 
suggesting an origin close to the central X-ray illuminating source.
We investigated time-resolved spectroscopy IC 4329a
to search for changes in the profile, fluxes, or peak energy of
the Fe K emission or 7.68 keV absorption features in IC 4329a, and explore if
any such variations are linked to continuum flux variations.

%%%%%%%%%%%%%   PN:   1st half versus second half

We started with short-term data and split the pn data into 
two time halves, each covering a duration of 
66.3 ksec and with a good time exposure of 
46.4 ksec. We fit each spectrum over 2.5--12 keV
with Model 12 (dual-diskline plus narrow Gaussian in absorption).
The best-fit spectral parameters, listed in Table 9, are all
consistent with each other;
differences in best-fit values can likely be attributed to
less-than-optimal photon statistics.
There is no evidence for temporal
evolution in any of the absorption and emission profile parameters.
The data/model residuals when each spectrum is fitted with a 
simple absorbed power-law are plotted in Figure 9 
and show no obvious changes in the profile.
However, given that the hard X-ray continuum 
light curve (shown in S05) displays minimal variability on these time scales,
this is not surprising. Additionally, given the minimal continuum variability,
we did not investigate spectral variability as a function of flux, i.e.,
high/low-flux states.

%%%%%%%%%

Next, we performed high time-resolution spectroscopy on all 
time scales to yield
light curves of $\Gamma$ and the Fe K flux.
Time bins were chosen with the goals of achieving adequate signal-to-noise
in the Fe K flux light curve.
The medium-term data were divided into 
18 bins, each covering a duration of 1.972 days,
or once every 30 satellite orbits;
each bin contained $\sim$10 separate {\it RXTE} visits.
The average exposure time per bin was 5.6 ksec.
The long-term data were divided into 18 bins
of approximately 44 days each in duration, with no bin overlapping
the two 56-day monitoring gaps:
data from the periods 2003 Apr 8 to 2003 Sep 29,
2003 Nov 24 to 2004 Sep 30,
and 2004 Nov 25 to 2005 Oct 2 were divided into four, seven, and seven bins, 
respectively. The average good time per bin was 8.1 ksec, excluding
bins 3 (100.9 ksec) and 4 (68.7 ksec), which overlapped with the
medium-term sampling.

Models were applied to time-resolved data based on the time-averaged fits.
The short-term data were divided into eight time bins,
with each bin covering a duration of 16.6 ksec
and with a good time exposure of 11.6 ksec.
Spectral fitting was done over 2.5--12 keV.
Model 5 (see Table 2), which featured 
Gaussians for the Fe K$\alpha$ and K$\beta$ emission lines, was used
for the short-term spectra.
Using the dual-diskline model offered no improvement to the
fits; also, the 7.68 keV absorption line was not detected in
any \xte\ spectrum. For both the long- and medium-term data, 
Model 3 (see Table 1), which featured a single Gaussian to model the 
Fe K$\alpha$ emission, was used. More detailed fits were not practical, 
e.g., the Fe K$\beta$ line was not detected. Spectral fitting was done 
over 3.5--24 keV. The HEXTE statistical errors were too large to 
study variability, so HEXTE was excluded from the time-resolved analysis.
The Fe K$\alpha$ line detected in nearly every time bin
at 99.8$\%$ confidence or greater in an F-test.
When fitting the time-averaged models, it was not practical to thaw
parameters such as the Fe K line profile shape or peak energy,
given the PCA resolution and sensitivity.
The neutral absorbing column density was kept as a free parameter.
The F-test was used to determine which free parameters to include in 
the fits. The time-resolved spectral fits were repeated
with each of photon index $\Gamma$, reflection fraction $R$,
and Fe K$\alpha$ line normalization $I_{\rm Fe}$ frozen at their
time-averaged values, and the values of total $\chi^2$ were compared
via an F-test to determine which parameters could justifiably allowed
to be free in the fits. The results are shown in Table 10.
We found no formal requirement to thaw $R$ on any time scale: the total
values of $\chi^2$ decreased only by 0.06 or 0.01
(long and medium, respectively)
when $R$ was thawed from the time-averaged value. 
Moreover, when much larger time bins were used to adequately constrain $R$,
we found $R$ to correlate strongly with $\Gamma$, but as noted by Vaughan \& Edelson (2001),
it is not possible to use the PCA to simultaneously constrain both $R$ and $\Gamma$
in an unbiased fashion, and false correlations can arise.
All analysis hereafter assumes $R$ is kept fixed at the
time-averaged value for each time scale. 
Strictly speaking, 
$R$ $\times$ continuum flux should be kept constant
in the light of a nonvariable Fe line flux (see below),
since both components are expected to have a common origin,
but for the current data this does not make any perceptible difference
in the fits.
As shown in Table 10, it was significant at 
$>$99.99 confidence to thaw $\Gamma$ only on the
medium and long time scales, suggesting statistically
significant variations in $\Gamma$ on these time scales.
It was significant at $>$99$\%$ confidence to thaw $I_{\rm Fe}$
only on the long time scale.
Errors on the Fe K line flux and $\Gamma$ light curves
were determined using the point-to-point variance,
as discussed in detail by Vaughan \& Edelson (2001) and Markowitz,
Edelson \& Vaughan (2003).
The 2--10 keV flux $F_{2-10}$ was measured from each model fit.
Errors on $F_{2-10}$ within a time bin were derived
from the 2--10 keV continuum light curves, using the
distribution of points within that time bin.

Mean spectral fit values and errors are listed in Table 11.
Figure 10 shows the light curves for $F_{2-10}$, $\Gamma$, and
$I_{\rm Fe}$ for all three time scales
We note that the long-term averages to the spectral fit parameters
shown in Figure 10 and listed in Table 11 do precisely match 
the spectral-fit parameters to the long-term time average-spectrum
listed in Table 1 (see $\S$4.4 for notes comparing 
the PCA and pn 2--10 keV relative flux normalizations).
This is due to the fact that the long-term time-averaged PCA spectrum
has a substantial contribution from the medium-term time-averaged spectrum
(the medium-term time-averaged spectrum comprises 36$\%$ of the long-term
spectrum's total exposure time).
Figure 11 shows correlation diagrams for $\Gamma$ and
$I_{\rm Fe}$ plotted against $F_{2-10}$. 
Table 11 also lists the Pearson correlation
coefficients and null hypothesis probabilities.
Finally, the fractional variability amplitudes, $F_{\rm var}$,
as defined in Vaughan \et\ (2003), are listed
in Table 12 for the light curves.

From Figures 10 and 11 and Tables 11 and 12, one can see
that there is not much variability in
$F_{2-10}$ or $\Gamma$ on short time scales.
However, $\Gamma$ tends to correlate well with $F_{2-10}$
on all three time scales, despite the 
somewhat narrow range spanned by $F_{2-10}$.
However, there is no strong evidence for variability in
$I_{\rm Fe}$ on any time scale; null hypothesis
probabilities for a constant line flux
cannot be rejected at greater
than 60$\%$ confidence on any time scale.
The lack of variability of the Fe line flux on
short time scales has already been reported
by S05. The results on long time scales are similar to those of
Weaver, Gelbord \& Yaqoob (2001), whose did not find any evidence 
for strong Fe line flux variability in IC~4329a based on five 
separate {\it ASCA} observations between 1993 and 1997.

\section{Discussion}

The \xmm\ pn spectrum of IC~4329a reveals an Fe K emission profile with 
two peaks, identified as Fe K$\alpha$ at 6.4 keV and Fe K$\beta$ at 7.0 keV. 
Modeling the 6.4 keV peak with a simple Gaussian reveals 
excess emission on the red side of the peak. The red wing can be fit either by
a Compton shoulder of $EW$ $\sim$ 10 eV, or by using
relativistic diskline models. $\S$6.1 discusses the emission profile in
the framework of relativistic diskline models.
The pn spectrum also reveals strong evidence for a narrow absorption feature
at 7.68 keV, likely due to H-like Fe K$\alpha$ 
blueshifted by 0.1$c$ relative to systemic,
suggesting a highly ionized outflow
of some sort. $\S$6.2 discusses the nature of this outflow.

\subsection{The Fe K Emission Profile: Moderately Relativistic Disklines}

The 6.4 keV peak is much stronger in terms of both $EW$ and height above 
the power-law continuum than the 7.0 keV peak, ruling out models wherein 
both peaks are the Doppler horns of a single relativistic diskline model 
(e.g., McKernan \& Yaqoob 2004). The red peak in particular is not symmetric: 
a moderate red wing is present, and models incorporating two symmetric peaks 
do not provide as good a fit to the data as models incorporating at least 
moderately-relativistic diskline emission from Fe K$\alpha$ and 
Fe K$\beta$. Alternatively, a dual-Gaussian fit plus 
a Compton shoulder provides an equally acceptable fit.
However, if the redward asymmetry is indeed due
to a diskline and not a Compton shoulder,
than IC~4329a's Fe K emission profile is thus distinct from 
more symmetric and more narrow profiles which lack a red wing, 
such as that seen in NGC 5548 (Yaqoob \et\ 2001, Pounds \et\ 2003a). 
Such narrow, symmetric profiles are generally suspected of having an 
origin in matter far from the black hole, such as in the outer 
accretion disk, broad line region, or the molecular torus. In IC~4329a, 
however, the presence of a moderate red wing leads us to conclude that
the emission profile is instead modeled equally well with 
the following two diskline scenarios: (1) each peak is modeled by a 
low-inclination, moderately broadened diskline, plus
little or no contribution from a narrow emission line, 
or (2) each peak is modeled
as the sum of a narrow line, to account for the emission 
closest to the rest energy of the line, plus a high-inclination 
diskline. In the latter case, neither the narrow
nor broad component dominates the emission.

In some Seyfert galaxies, identification of broad Fe lines is complicated by model
degeneracies which arise when 3--6 keV spectral curvature due to warm 
absorption is taken into account (e.g., Reeves \et\ 2004, Turner \et\ 2005).
In IC~4329a, however, only minimal spectral curvature above 3 keV, and no 
spectral curvature above 5 keV, is expected given the results of soft 
X-ray absorption modeling (see Figure 7b),
and so the detection of a moderate red wing is robust. The best-fitting 
model includes emission extending in to a radius of 
$<$26 $R_{\rm g}$, similar to that found by Done \et\ (2000) 
using {\it ASCA}. Done \et\ (2000) note that IC~4329a's profile
does not seem as broadened and reddened as that of MCG--6-30-15, 
where the diskline emission extends down to $\sim$2--3$R_{\rm g}$ 
and the emissivity index is relatively steep, 
favoring emission from the central
regions of the disk (Wilms \et\ 2001; Young \et\ 2005). 
The red wing emission in IC~4329a 
extends down only to energies of $\sim$5.4 keV (Done \et\ 2000) or $\sim$5.7 keV
(this work).
Although our analysis cannot formally rule out an inner emission radius
of $\lesssim$2$R_{\rm g}$, the Fe K emission profile in IC~4329a thus seems 
to fall into a category between those consisting of narrow lines 
and those consisting of very strongly relativistically broadened lines.

%%%%%%%We find a FWHM of the K$\alpha$ peak of $\sim$10000 km s$^{-1}$. 
%%%%%%%%FOR CASE 1 only

%%%%%%%Assuming that the bulk of the Fe K emission originates in     
%%%%%%%the inner regions of an optically-thick, geometrically-thin accretion disk, 
%%%%%%%as in case 1) above, these findings are consistent with the notion that 
%%%%%%%the accretion disk in IC~4329a could be truncated below 
%%%%%%%$\sim$10--20 $R_{\rm g}$, or could transition from a geometrically-thin,
%%%%%%%optically-thick disk at large radii to a radiatively-inefficient flow, such
%%%%%%%as an advection-dominated accretion flow (ADAF; Narayan \& Yi 1995), 
%%%%%%%in the innermost regions of the accretion disk, e.g.,
%%%%%%%Dove \et\ (1997) and Esin, McClintock \& Narayan (1997).

The lack of variability in the Fe K emission line is consistent with an 
origin in distant matter, as suggested, e.g., by S05. However, the line
profile suggests a nonnegligible contribution from material
origin close to the black hole. If in fact the bulk of the Fe K
emission originates in the inner accretion disk, as in case (1),
then a likely scenario is 
as follows: assuming that variations in the line track those in the 
X-ray continuum, modified only by a light travel time delay, the line 
will not vary strongly unless the continuum does. However, the continuum 
in IC~4329a does not display strong variability. Even on time scales 
of $\sim$2 years, the peak-to-trough continuum variations do not exceed 
$\sim$50$\%$, which simply may not be sufficient to trigger very strong 
trends in Fe K flux. Similarly, with time bins chosen to maximize the 
variability-to-noise in the Fe flux light curves, 
this analysis may simply not be sufficiently sensitive to track 
small-amplitude trends ($<$50$\%$ variations) in line flux.

If there are approximately equal broad and narrow component contributions
to the profile, as in case 2), then, for time scales longer than the 
light-crossing times between the continuum source and the sites of
broad and narrow line production, 
it could again be the case that the lack of variability in either component
is a result of the lack of strong continuum variability. However,
the PCA is not sensitive enough to distinguish between the broad and narrow
components. Specifically, if the narrow component, which comprises $\sim$60$\%$
of the total line emission, is constant, then the PCA time-resolved
spectroscopy will not be sensitive to small changes in the broad component.
%%% For time scales shorter than the light crossing timescale
%%% for the narrow line   due to combination of effects:
%%%% continuum doesn't vary AND large constribution from distant material.
In any case, the lack of strongly correlated variations between the
continuum and line prevents us from making any firm conclusions
about using variability to determine the location of the line-emitting
material.

\subsection{The 7.68 keV Absorption Feature: The Highly Ionized Outflow
in IC~4329a}

High-resolution spectroscopy of the Fe K band of several AGN
has yielded evidence for highly ionized absorbing gas,
e.g., NGC 3783 (Reeves \et\ 2004) and MCG--6-30-15 (Young \et\ 2005).
In addition, there is evidence in PG and BAL quasars and one
other high-luminosity Seyfert for Fe K lines and/or edges
that suggest absorbing gas that is both highly ionized
and outflowing at relativistic velocities. For instance,
in PG~1211+143 (Pounds \et\ 2003b), PDS 456 (Reeves \et\ 2003),
the BAL quasars APM 08279+5255 (Chartas \et\ 2002; Hasinger,
Schartel \& Komossa 2002) and PG~1115+080 (Chartas \et\ 2003),
and the Seyfert galaxy Mkn 509 (Dadina \et\ 2005), the
ionization parameters were typically log$\xi$ $\sim$
2.5 -- 3.7, the outflow velocities spanned 0.08$c$--0.34$c$,
and the absorber column densities were typically
$\sim$1--5 $\times 10^{23}$ cm$^{-2}$.

%%%PG1211        24000 km/s     xi = 3.4    NH = 5E23
%%%PG0844        0.2c              3.7    NH=4E23
%%%APM082     0.2c and 0.4c      Fe 25       NH = 1E23
%%%PG1115     0.10c   0.34c      Fe 25      NH = 4E23
%%%PDS 456         50,000 km/s      xi=3.5   NH = 5E23

%%%% IC          0.09c          xi =3.7    NH = 1.4E22

The highly ionized (log$\xi$ $\sim$ 3.7), high-velocity ($\sim$0.1$c$ blueshift 
relative to systemic) absorbing material in
IC~4329a has a derived ionization parameter
and outflow velocity similar to what is found in these objects,
though we note that the derived column density for the
IC~4329a's outflow component is about
a factor of 10 lower. Nonetheless, we hereby add
IC~4329a to this rather small but growing list, noting that
IC~4329a is thus the lowest redshift AGN known to exhibit this type of outflow.

To estimate the distance $r$ between the central black hole
and the outflowing gas, we can use $\xi$ = $L_{2-200}$/($n$ $r^2$),
where $n$ is the number density.
$L_{2-200}$ is the 2--200 keV illuminating continuum luminosity.
%%%% , which we use instead of the 1--1000 Rydberg ionizing continuum luminosity 
%%%% because most of the photons that ionize this particular component are expected to be hard, not soft, X-rays.
We estimate the maximum possible
distance to the material by assuming that the thickness
$\Delta$$r$ must be less than the distance $r$.
The column density $N_{\rm H}$ = $n$$\Delta$$r$,
yielding the upper limit $r$ $<$ $L_{2-200}$/($N_{\rm H}$$\xi$).
%%%The ionizing luminosity, estimated by referring to Model 12 and removing
%%%all absorbing components, is 1.4$\times$10$^{44}$ erg s$^{-1}$.
We estimate the 2--200 keV flux from our models to be 
$4.9 \times 10^{-10}$ erg cm$^{-2}$ s$^{-1}$
(assuming $H_{\rm o}$ = 70 km s$^{-1}$ Mpc$^{-1}$ and $\Lambda_{\rm o}$ = 0.73), 
which corresponds to 
$L_{2-200} = 2.8 \times 10^{44}$ erg s$^{-1}$.
This yields $r < 3.8 \times 10^{18}$ cm, or 1.2 pc.

In the aforementioned PG and BAL quasars, the derived mass outflow rates
are usually at least 1 $\Msun$ yr$^{-1}$ and
on the order of the accretion inflow rate,
suggesting that these outflows account
for a substantial portion of the total energy budget of the AGN.
Under the assumption that the gas is in equilibrium and that
the outflow velocity is a constant, the mass outflow rate \.{M}$_{\rm out}$ for
the IC~4329a component can be derived via conservation of mass:
\.{M}$_{\rm out}$ = $\Omega$$n$$r^2$$v$$m_{\rm p}$,
where $v$ is the outflow velocity, $m_{\rm p}$ is the proton mass,
and $\Omega$ is the covering fraction, which we will assume
is 4$\pi$/10 sr.
Substituting $n$$r^2$ = $L_{2-200}$/$\xi$,
we find \.{M}$_{\rm out}$ $\sim 3 \times 10^{26}$ gm s$^{-1}$ = 
5 $\Msun$ yr$^{-1}$. 
We note that
the actual outflow rate should be lower if there is an extreme degree of
collimation, or higher if the outflowing component 
is not directed along the line of sight.

We can compare this to the inflow accretion rate \.{M}$_{\rm acc}$ using
$L_{\rm bol}$ = $\eta$\.{M}$_{\rm acc}$$c^2$,
where $\eta$ is the accretion efficiency parameter, typically 0.1.
Using the \xmm\ 
2--10 keV flux of $8.9 \times 10^{-11}$ erg cm$^{-2}$ s$^{-1}$, we derive 
an 2--10 keV luminosity $L_{2-10}$ of $5.1 \times 10^{43}$ erg s$^{-1}$. 
Using the relation of 
Padovani \& Rafanelli (1988), $L_{\rm bol}$ = 59$\nu$$L_{\nu}$ at 
$\nu$=2 keV, and a photon index $\Gamma$ of 1.75, the bolometric 
luminosity is thus estimated to be $L_{\rm bol}$ = 32$L_{2-10}$ = 
$1.6 \times 10^{45}$ erg s$^{-1}$. 
We find \.{M}$_{\rm acc}$ $\sim$ 0.3 $\Msun$ yr$^{-1}$.
The kinetic power associated with the outflow component,
estimated as \.{M}$_{\rm out}$$v^2$, is 3$\times$10$^{45}$ erg s$^{-1}$, 
roughly the same as the bolometric luminosity.
Similar to the situation in the aforementioned quasars,
the outflow rate of this particular absorbing component may
therefore represent a substantial fraction
of the accretion inflow used to power the bolometric luminosity in IC~4329a.
Similar calculations on the low-velocity,
soft X-ray absorbing components (S05), using the 1--1000 Rydberg
ionizing continuum luminosity $L_{\rm ion}$ 
estimated to be $1.4 \times 10^{44}$ erg s$^{-1}$,
suggest that they, too,
are associated with outflow rates that are of the same magnitude or higher
than the inflow accretion rate. 
Our estimate for \.{M}$_{\rm out}$ is a rough figure, 
but the fact that it is an order of magnitude higher than \.{M}$_{\rm acc}$
suggests that if the outflow were sustained for very long periods, 
matter could not be transported into the black hole
and it could not grow. A low covering fraction 
and/or a non-continuous, intermittent outflow is required to 
simultaneously have an outflow existing and the black hole
being fed over time.

%%%%%%%%%%%%%%%%%%%%%%%

The high velocity of this outflow suggests that it may  
be associated with the accretion disk at a small radius. 
An explanation commonly invoked for the high-velocity, high-ionization
absorbers in that the flow originates in the innermost part of
a radiatively driven accretion disk wind (e.g., Proga, Stone \& 
Kallman 2001; King \& Pounds 2003), although one caution is that it 
is generally difficult to accelerate highly ionized ions using 
radiation pressure alone. In this scenario, the winds 
must be launched from very close in to the black hole, yet far 
enough away from the black hole that the outflow velocity exceeds 
the local escape velocity. For a Keplerian disk, the radius 
at which the escape velocity is 0.1$c$ will be equal to 
($v_{\rm esc}$/$c$)$^2$$R_{\rm Sch}$ = 100$R_{\rm Sch}$;
a radiatively driven wind could thus be applicable to IC 4329a
if (neglecting acceleration) the wind is launched from
a larger radius.   

More specifically, 
King \& Pounds (2003; see also Reeves \et\ 2003) demonstrated
that objects accreting near the Eddington rate
are likely to exhibit radiatively driven winds with substantial column densities,
and the outflow rates of these winds may be comparable to the accretion inflow rate.
If the flow becomes optically-thick at small radii
($\sim$10--100 $R_{\rm g}$), extreme velocities and high column densities are 
likely to the associated with the flow.
This scenario may be applicable to the high-velocity absorber in IC 4329a.
However, we do not know {\it a priori} if the accretion rate of IC 4329a
is indeed near Eddington.
Uncertainty regarding IC 4329a's black hole mass $M_{\rm BH}$ prevents
precise knowledge about the accretion rate. 
The reverberation-mapped estimate for
$M_{\rm BH}$ (Peterson \et\ 2004) is formally an upper limit only,
$9.9^{+17.9}_{-9.9} \times 10^6 \Msun$, 
implying an accretion rate relative to 
Eddington of $L_{\rm bol}$/$L_{\rm Edd}$ of $>$0.5 (best estimate, 1.29).
Other methods yield higher black hole mass estimates. For example,
using the photoionization method (estimating velocity dispersions based on
H$\beta$ line widths) to estimate the distance from the central illuminating
source to the broad-line region, Wandel, Peterson \& Malkan (1999) 
estimate a black hole mass of $2.2 \times 10^7 \Msun$. Along another track,
Nikolajuk, Papadakis \& Czerny (2004) suggest a prescription to estimate 
$M_{\rm BH}$ on the basis of short-term X-ray variability amplitude measurement, 
specifically the normalized excess variance, $\sigma^2_{\rm NXS}$. They 
estimate $M_{\rm BH}$ = $1.24 \times 10^8 \Msun$ based on short-term 
\xte\ light curves. We rebinned the short-term, pn 2.5--12 keV light curve 
with a sampling time of 6.8 ksec, to yield a light curve with enough data 
points (20) to get an accurate measurement of $\sigma^2_{\rm NXS}$. We found 
$\sigma^2_{\rm NXS}$ = 0.00136 $\pm$ 0.00010 (errors derived using Vaughan \et\ 
2003); the Nikolajuk \et\ (2004) relation yields a $M_{\rm BH}$ estimate 
of $8.64 \pm 0.60 \times 10^{7} \Msun$, similar to the Nikolajuk \et\ 
(2004) estimate\footnote{We note that the Nikolajuk \et\ (2004) method 
assumes that the power spectral density function (PSD) contains a
'break' (change in power-law slope from --2 to --1) at temporal
frequencies corresponding to time scales larger than the duration over
which one is measuring $\sigma^2_{NXS}$ (see Nikolajuk \et\ 2004 for 
derivation, as well as additional assumptions and caveats). PDS measurement 
for IC~4329a is still in progress, pending accumulation of long-term 
monitoring data, (A.\ Markowitz et al., in prep.), but we will assume that the 
PSD break for IC~4329a corresponds a time scale longer than 136 ksec (as 
expected for any Seyfert with a black hole mass larger than 
$5 \times 10^6 \Msun$; see Markowitz \et\ 2003).}.
This value of the black hole mass places the estimated accretion rate
$L_{\rm bol}$/$L_{\rm Edd}$ at 15 $\pm$ 1 $\%$.
A more accurate black hole mass determination, e.g., via 
more accurate reverberation mapping, is needed to resolve this issue.
We can only conclude that we currently cannot rule out an accretion rate
close to the Eddington limit, and so a radiatively-driven disk wind 
cannot be ruled out.

%%%%%%%%%%%%

Another problem with the disk-wind scenario is that the 7.68 keV
absorption feature is narrow, and consistent with a single
component only. A broad range of velocities seems more physically 
plausible if we are seeing a continuous stream of gas being accelerated 
from rest to 0.1$c$ (relative to the systemic velocity). However, there are 
no such indications in the Fe K bandpass of material at lower velocities.
These facts suggest that we could be witnessing a discrete, transient, 
outflowing blob of material as opposed to a continuous flow.

In this case, the average mass outflow rate will be lower
than the estimate given above, depending on the duty cycle of ejection.
One constraint to note is that 
there are no indications from the soft X-rays
(e.g., S05) for absorption at a similar relativistic
outflow velocity, so the absorbing material
would have to be in the form of a single blob consisting of 
a high-ionization component only, with no
lowly- or moderately ionized contribution, 
for such a scenario to be applicable.

In addition to radiatively driven outflows, another mechanism whereby 
radio-quiet AGN can have material outflowing at relativistic velocities 
is described by the so-called aborted-jet model of Ghisellini, 
Haardt \& Matt (2004). In this model, radio-quiet accreting black 
holes launch jets only intermittently, yielding discrete, fast-moving blobs
that travel along the black hole rotation axis, as opposed to a jet that 
is continuously on. Extraction of black hole rotational energy, in addition to 
accretion energy, provides the energy source.
The jet is launched at a velocity less the escape velocity, so blobs
reach a maximum radius from the black hole before falling backwards
and colliding with later-produced outflowing blobs.
In the case of IC~4329a, we could thus be witnessing a singular, transient
blob at a particular point in its outward journey, although its location
must be $<$100 $R_{\rm Sch}$ for this scenario to apply.

Finally, we note that this high-velocity hard X-ray absorber
is unlikely to be directly physically connected
to the soft X-ray absorbers seen by S05. 
The soft X-ray absorbers are likely too low in outflow velocity,
column density, and ionization state 
to be part of the same outflow phenomenon as the hard X-ray absorber,
suggesting physically distinct phenomena.
There is no evidence that the soft X-ray absorbing material ``feeds''
or is ``fed by'' the hard X-ray absorber, and no obvious 
acceleration/deceleration mechanism.

\section{Conclusions}

We have presented a detailed analysis of the \xmm\ pn long-look
spectrum of IC 4329a. The Fe K bandpass is 
dominated by two peaks, identified as neutral
or near-neutral Fe K$\alpha$ and K$\beta$ emission.
Both peaks are resolved; the K$\alpha$ peak is estimated to
have a FWHM velocity near 10,000 km s$^{-1}$.
There is excess emission on the red side of the 6.4 keV line,
near 6.1 keV, 
and we find that three models can all describe the overall
emission profile approximately equally well:
(1) a model in which the Fe K$\alpha$ and K$\beta$ lines are
described by Gaussians, signifying neutral or near-neutral Fe emission,
and the $\sim$6.1 keV emission is a Compton shoulder;
(2) a model in which
low-inclination dual-diskline profiles originating in mildly ionized Fe dominate
the peaks, with little or no contribution
from a narrow component; or (3) a model in which approximately
equal contributions from a high-inclination diskline
and a narrow component describe each peak.
The resolution of an X-ray calorimeter is needed
to deconvolve the narrow and broad components
and break the degeneracy between these various cases.
However, the 6.4 keV peak is much higher in equivalent width and height above
the continuum compared to the 7.0 keV peak, ruling out models
whereby both peaks are the Doppler horns of a single relativistic
diskline. 
There is no strong evidence for emission or absorption at 
the systemic velocity by \ion{Fe}{25} or \ion{Fe}{26}.
In addition, we have used \xte\ monitoring data to 
extract a PCA + HEXTE spectrum to derive tight
constraints on the strength of the reflection component. 

We have performed time-resolved spectral fitting on the \xmm\ long-look
and the \xte\ monitoring data
to probe the variability of the X-ray continuum, photon index
and Fe K line flux on timescales spanning $\lesssim$a day
to 2 years. The photon index and 2--10 keV flux are well correlated.
However, there is no strong evidence for variability in the Fe K line
on any time scale probed, likely due to the minimal level of continuum 
variability displayed.

We find strong evidence for a narrow absorption line at 7.68 keV.
We have used the Monte Carlo simulation method outlined 
in Porquet \et\ (2004) to confirm the high level of
detection significance and constrain the probability that
the feature is due to photon noise to $<$0.1$\%$.
The Monte Carlo method, as used here, 
accounts for uncertainty in the model, a minor improvement
compared to Porquet \et\ (2004). 
The most likely origin for the line is \ion{Fe}{26},
in highly ionized material 
blueshifted by $\sim$0.1$c$ relative to systemic, and
with a column $\sim 1 \times 10^{22}$ cm$^{-2}$.
This phenomenon is similar to what is seen
in other high luminosity AGN, including PG and BAL quasars,
making IC 4329a the lowest redshift AGN known to exhibit this
phenomenon. The mass outflow rate of this component is estimated 
to a few $\Msun$ yr$^{-1}$, and the associated kinetic power is
estimated to be of the same order as the bolometric luminosity, 
suggesting that the outflow represents at least a  
substantial portion of the overall energy budget.
The outflow may be associated with a radiatively driven disk wind.
However, because of the narrow velocity range seen, it is also possible 
that the absorbing material is in the form of a discrete blob
of emission. It is important to find as many of these
absorbing features as possible in AGN spectra in order
to correctly gauge their frequency of occurrence.

\acknowledgements

The authors thank Katrien Steenbrugge for insightful comments 
and suggestions, for providing the model used to make Figure 7b, and
for providing early access to EPIC proprietary data.
The authors thank Philip Uttley for providing his
Monte Carlo spectral simulation code.
A.M.\ thanks Nikolai Shaposhnikov, Jean Swank, and Craig Markwardt for
guidance on HEXTE data reduction and spectral analysis.
The authors also thank the referee for useful comments.
This work has made use of observations obtained with {\it XMM-Newton}, an 
ESA science mission with instruments and contributions directly funded by 
ESA member states and the US (NASA). 
This work has made use of data obtained through the High Energy
Astrophysics Science Archive Research Center Online Service, provided by
the NASA Goddard Space Flight Center, and the NASA$/$IPAC Extragalactic 
Database which is operated by the Jet Propulsion Laboratory, California 
Institute of Technology, under contract with the National Aeronautics 
and Space Administration.

\clearpage

%%%%%%%%Table 1   Joint PCA/HEXTE fitting results for Model 3

\begin{deluxetable}{lcc}
\tabletypesize{\footnotesize}
\tablewidth{6.5in}
\tablenum{1}
\tablecaption{Joint PCA/HEXTE fitting results for Model 3: power-law + {\sc pexrav} + Gaussian\label{tab1}}
\tablehead{
\colhead{Parameter}                        &   \colhead{Long Term}                 &    \colhead{Medium Term}      }
\startdata
$\chi^2/dof$                               & 208.14/123                                 & 82.7/101       \\
2--10 keV flux   (erg cm$^{-2}$ s$^{-1}$)  &  1.20 $\times$ 10$^{-10}$                    &  1.12 $\times$ 10$^{-10}$  \\
25--100 keV flux (erg cm$^{-2}$ s$^{-1}$)  &  1.78 $\times$ 10$^{-10}$                    &  1.74 $\times$ 10$^{-10}$  \\
$\Gamma$                                   & 1.894$^{+0.013}_{-0.015}$                  & 1.782$^{+0.012}_{-0.021}$      \\
$R$                                        & 0.51 $\pm$ 0.04                              & 0.35$^{+0.05}_{-0.11}$          \\
$E$ (keV)                                  & 6.38$^{+0.05}_{-0.02}$                     & 6.33 $\pm$ 0.03 \\ 
$\sigma$ (eV)                              &  228 $\pm$ 50                              & 341 $\pm$ 54          \\
$I$ (10$^{-5}$ ph cm$^{-2}$ s$^{-1}$)     &  18.2$^{+1.4}_{-1.1}$                       & 25.3 $\pm$ 1.9 \\
$EW$  (eV)                                   & 128$^{+10}_{-8}$                           & 197 $\pm$ 15           \\
$N_H$ (10$^{20}$ cm$^{-2}$)                & 90 $\pm$ 20                                  & $<$40          \\
\enddata
\tablecomments{Model parameters for the best-fitting Model 3 (power-law + {\sc pexrav} + Gaussian)                 
to the PCA/HEXTE data. Errors are 90$\%$ confidence for one interesting parameter. 
$E$, $\sigma$, $K$, and $EW$ are the centroid energy, width, intensity and equivalent width for
the Fe K$\alpha$ Gaussian emission component.
Fe K$\beta$ line emission was not considered here.
The high-energy cutoff in the {\sc pexrav} component was fixed at 270 keV (e.g., Perola \et\ 1999);
the inclination was kept fixed at 30$\degr$.
The 25--100 keV flux is based on average of HEXTE cluster A and B model fluxes.}
\end{deluxetable}

%%%%$A$            & 4.34$^{+0.11}_{-0.13}$ $\times$ 10$^{-2}$  &      3.18 $\pm$ 0.04 $\times$ 10$^{-2}$          \\  %%%verify

%%%%%%%%%%%%%%%%%%%%%%%%%%%%%%%%%%%%%%%%%%%%%%
\begin{deluxetable}{llccccc}
\tabletypesize{\footnotesize}
\tablewidth{6.5in}
\tablenum{2}
\tablecaption{Initial EPIC-pn model fits for Fe K emission lines\label{tab2}}
\tablehead{
\colhead{}&\colhead{}&\colhead{Model 1}&\colhead{Model 2}&\colhead{Model 3}&\colhead{Model 4} &\colhead{Model 5}}  
\startdata
\multicolumn{2}{l}{$\chi^2/dof$}      & 2598.24/1694                         
                                      & 1727.57/1691  
                                      & 1641.80/1691 
                                      & 1622.47/1688  
                                      & 1630.04/1691 \\
\multicolumn{2}{l}{$\Gamma$}          & 1.656$^{+0.015}_{-0.011}$            
                                      & 1.671$^{+0.012}_{-0.015}$            
                                      & 1.737 $\pm$ 0.012 
                                      & 1.743 $\pm$ 0.013  
                                      & 1.742 $\pm$ 0.013  \\
\multicolumn{2}{l}{$N_H$ (10$^{20}$ cm$^{-2}$)}   &  $<$10
                                                  &  $<$7 
                                                  &  $<$16
                                                  &  $<$17
                                                  &  $<$17 \\
Red peak & E (keV)                    & ... & 6.40 $\pm$ 0.01 
                                            & 6.40 $\pm$ 0.01 
                                            & 6.39 $\pm$ 0.01 
                                            & 6.40 $\pm$ 0.01 \\
         & $\sigma_{6.4}$                 & ... & 106 $\pm$ 14 
                                            & 91 $\pm$ 15  
                                            & 91 $\pm$ 13 
                                            & 93$^{+15}_{-13}$ \\
 &$I_{\rm K\alpha}$     & ... & 8.8 $\pm$ 0.7  
                                            & 7.8$^{+0.4}_{-0.6}$ 
                                            & 8.0$^{+0.6}_{-0.7}$  
                                            & 8.0 $\pm$ 0.6    \\
         & $EW_{\rm K\alpha}$ (eV)              & ... & 86 $\pm$ 6 
                                            & 75$^{+4}_{-6}$ 
                                            & 76$^{+6}_{-7}$  
                                            & 77 $\pm$ 6 \\
Blue peak & E (keV)                   & ... & ... & ... & 6.93$^{+0.07}_{-0.09}$  &  7.06 (fixed)\\
         & $\sigma_{7.0}$                 & ... & ... & ... & 121$^{+95}_{-59}$       &  (= $\sigma_{6.4}$) \\
& $I_{\rm K\beta}$ & ... & ... & ... & 1.3$^{+0.7}_{-0.5}$     &  (= 0.13 * $I_{K\alpha}$)\\
         & $EW_{\rm K\beta}$ (eV)         & ... & ... & ... & 15$^{+8}_{-6}$          &  (= 0.13 * $EW_{K\alpha}$) \\
\enddata
\tablecomments{Best-fit spectral parameters for various models.
Models 4 and 5 additionally included a Gaussian 
to model Fe K$\beta$ emission with centroid energy free and fixed, respectively. 
Line intensities are in units of 10$^{-5}$ ph cm$^{-2}$ s$^{-1}$.
A {\sc pexrav} component was included in all models except 1 and 2, with reflection strength $R$ fixed at 0.51.  
The 2--10 keV flux for Models 4--5 was $8.9 \times 10^{-11}$ erg cm$^{-2}$ s$^{-1}$. See text for additional details.}
\end{deluxetable}

\begin{deluxetable}{llc}
\tabletypesize{\footnotesize}
\tablewidth{6.5in}
\tablenum{3}
\tablecaption{EPIC-pn model fits: Compton shoulder emission (Model 6) \label{tab3}}
\tablehead{
\colhead{}&\colhead{}&\colhead{}}
\startdata
\multicolumn{2}{l}{$\chi^2/dof$}                  & 1611.72/1686    \\
\multicolumn{2}{l}{$\Gamma$}                      & 1.744 $\pm$ 0.012     \\
\multicolumn{2}{l}{$N_H$ (10$^{20}$ cm$^{-2}$)}   & 9$^{+7}_{-5}$  \\  
Red peak & E (keV)                                & 6.40 $\pm$ 0.01  \\ 
         & $\sigma_{6.4}$                             & 76$^{+24}_{-19}$  \\
 &$I_{K\alpha}$ (10$^{-5}$ ph cm$^{-2}$ s$^{-1}$) & 7.4$^{+1.1}_{-1.0}$  \\
         & $EW_{\rm K\alpha}$ (eV)                    & 70$^{+11}_{-9}$ \\
Blue peak & E (keV)                               & 7.06 (fixed)\\
         & $\sigma_{7.0}$                             & (= $\sigma_{6.4}$)  \\
& $I_{K\beta}$ (10$^{-5}$ ph cm$^{-2}$ s$^{-1}$)  & (= 0.13 * $I_{K\alpha}$) \\
         & $EW_{\rm K\beta}$ (eV)                     & (= 0.13 * $EW_{\rm K\alpha}$)  \\
Compton shoulder &  E (keV)                       & 6.13$^{+0.07}_{-0.25}$ \\
         & $\sigma_{7.0}$                             & (= $\sigma_{6.4}$)  \\
& $I$  (10$^{-5}$ ph cm$^{-2}$ s$^{-1}$)          & 1.1 $\pm$ 0.5 \\
         & $EW$ (eV)                                & 9 $\pm$ 5 \\
\enddata
\tablecomments{Best-fit spectral parameters to a model with Gaussians for Fe K$\alpha$, Fe K$\beta$,
and Compton shoulder emission.}
\end{deluxetable}

\begin{deluxetable}{lc}
\tabletypesize{\footnotesize}
\tablewidth{6.5in}
\tablenum{4}
\tablecaption{EPIC-pn model fits using a single diskline component (Model 7) \label{tab4}}
\tablehead{  \colhead{}&\colhead{ }} 
\startdata
$\chi^2/dof$                       & 1705.56/1691 \\
$\Gamma$                           & 1.739 $\pm$ 0.013 \\
$N_H$ (10$^{20}$ cm$^{-2}$)        &   8$^{+7}_{-4}$            \\  
$E$ (keV)                          & 6.40 $\pm$ 0.01 \\
$I$ (10$^{-5}$ ph cm$^{-2}$ s$^{-1}$)   &  8.6$^{+0.5}_{-0.6}$ \\
$EW$                                    &   84$^{+6}_{-6}$ \\
$\beta$                       & 6 (unconstrained)    \\
$R_{\rm in}$                      & 300$^{+100}_{-270}$  \\ 
$i$                        & 30$\degr$ (fixed) \\
\enddata
\tablecomments{The outer radius was fixed at
400 $R_{\rm g}$. Results are presented here for an inclination angle $i$ of 30$\degr$; angles of
50$\degr$ or 70$\degr$ yielded significantly worse fits.}
\end{deluxetable}

\begin{deluxetable}{llc}
\tabletypesize{\footnotesize}
\tablewidth{6.5in}
\tablenum{5}
\tablecaption{EPIC-pn model fits using dual diskline components (Model 8) \label{tab5}}
\tablehead{  \colhead{}&\colhead{ }} 
\startdata
\multicolumn{2}{l}{$\chi^2/dof$}                  & 1612.65/1688 \\
\multicolumn{2}{l}{$\Gamma$}                      & 1.741 $\pm$ 0.013 \\
\multicolumn{2}{l}{$N_H$ (10$^{20}$ cm$^{-2}$)}   & $<$14           \\  
Red peak         & $E$ (keV)                      & 6.44 $\pm$ 0.01 \\
                 & $I$ (10$^{-5}$ ph cm$^{-2}$ s$^{-1}$)  &  8.9$^{+1.7}_{-0.8}$ \\
                 & $EW$                           &   86$^{+17}_{-8}$ \\
Blue peak        & $E$ (keV)                      & 6.98 $\pm$ 0.09 \\
\multicolumn{2}{l}{$\beta$}                       & 2.1 $\pm$ 0.3     \\
\multicolumn{2}{l}{$R_{\rm in}$}                      & $<$26  \\ 
\multicolumn{2}{l}{$i$  }                       & $<$12$\degr$ \\
\enddata
\tablecomments{For each {\sc Laor} diskline component, the outer radius was kept fixed at
400 $R_{\rm g}$. The normalization of the blue peak diskline was fixed at 0.13 times that
of the red peak.}
\end{deluxetable}

\begin{deluxetable}{lllcc}
\tabletypesize{\footnotesize}
\tablewidth{6.5in}
\tablenum{6}
\tablecaption{EPIC-pn model fits using dual-disklines and dual-Gaussians \label{tab6}}
\tablehead{ \colhead{}& \colhead{}&\colhead{ } &\colhead{Model 9} &\colhead{Model 10}    }
\startdata
\multicolumn{3}{l}{$\chi^2/dof$}                                   &  1610.21 / 1687           & 1614.06 / 1686                \\
\multicolumn{3}{l}{$\Gamma$}                                       &  1.741 $\pm$ 0.013        & 1.743 $\pm$ 0.013               \\
\multicolumn{3}{l}{$N_H$ (10$^{20}$ cm$^{-2}$)}                    &  $<$14                    & $<$14  \\ 
Gaussians            & Red peak   & $E$ (keV)                      &  6.40 (fixed)             & 6.39 $\pm$ 0.01      \\
                     &            & $I_{\rm K\alpha,narrow}$       &  $<$5.9                   & $<$7.2   \\
                     &            & $EW_{\rm K\alpha,narrow}$ (eV) &  $<$52                    &  $<$66   \\
                     & Blue peak  &  $E$ (keV)                     &  7.06 (fixed)             &  7.06 (fixed)     \\
                     & \multicolumn{2}{l}{$\sigma$}                &  10 (fixed)               &  66$^{+19}_{-22}$  \\ 
{\sc Laor} Disklines & Red peak   & $E$ (keV)                      &  6.42$^{+0.19}_{-0.15}$   &  6.30$^{+0.40}_{-0.22}$ \\
                     &            & $I_{\rm K\alpha,broad}$        &  6.9 $\pm$ 3.0            &  4.0 $\pm$ 1.8    \\
                     &            & $EW_{\rm K\alpha,broad}$ (eV)  &  65 $\pm$ 28              &  37 $\pm $17      \\
                     & Blue peak  & $E$ (keV)                      & 6.83$^{+0.12}_{-0.41}$    &  7.06 (fixed)         \\
                     & \multicolumn{2}{l}{$\beta$}                 & 2.5$^{+1.4}_{-0.8}$       &  1.8$^{+2.4}_{-1.6}$  \\
                     & \multicolumn{2}{l}{$R_{\rm in}$}            & $<$64                     &  $<$81                \\ 
                     & \multicolumn{2}{l}{$i$}                     & $<$30$\degr$              &  45$^{+19}_{-41}$$\degr$ \\
\enddata
\tablecomments{Best-fit spectral parameters for a model
in which the red peak (Fe K$\alpha$) and blue peak (Fe K$\beta$) 
are each modeled by the sum of a narrow Gaussian and a broad {\sc Laor} diskline.
The widths $\sigma$ of the two Gaussians were tied.
The intensity of the blue peak Gaussian was fixed at 0.13
that of the red peak Gaussian. The inner radii $R_{\rm in}$,
emissivity index $\beta$ and inclination $i$ of the two diskline
components were tied. The outer radii of both disklines were
kept fixed at 400 $R_{\rm g}$. 
Line intensities are in units of 10$^{-5}$ ph cm$^{-2}$ s$^{-1}$.
The intensity of the blue peak diskline was kept fixed at 0.13
that of the red peak diskline. To get a reasonable fit, the centroid energy
of the K$\beta$ Gaussian and the rest energy of the K$\beta$ diskline
were both kept fixed at 7.06 keV. A {\sc pexrav} component with $R$ = 0.51
was included in the model.} %%%%%  lower value Eo Kbeta, model 9a pegged
\end{deluxetable}

\begin{deluxetable}{llcccc}
\tabletypesize{\footnotesize}
\tablewidth{6.5in}
\tablenum{7}
\tablecaption{EPIC-pn model fits for 7.68 keV absorption feature\label{tab7}}
\tablehead{
\colhead{}&\colhead{}&\colhead{Model 11}&\colhead{Model 12}&\colhead{Model 13} }  
\startdata
\multicolumn{2}{l}{$\chi^2/dof$}      & 1605.81/1686  
                                      & 1589.46/1685 
                                      & 1584.92/1685 \\
\multicolumn{2}{l}{$\Gamma$}          & 1.715$^{+0.014}_{-0.006}$            
                                      & 1.734 $\pm$ 0.012 
                                      & 1.728$^{+0.014}_{-0.009}$  \\
\multicolumn{2}{l}{$N_H$ (10$^{20}$ cm$^{-2}$)}   &  $<$12
                                                  &  $<$12
                                                  &  $<$10  \\
Red peak & $E$ (keV)        & 6.44 $\pm$ 0.01 & 6.44 $\pm$ 0.01 & 6.44 $\pm$ 0.01 \\
         & $I$ (10$^{-5}$ ph cm$^{-2}$ s$^{-1}$)   &  8.0 $\pm$ 0.7
                                                   &  8.5$^{+0.9}_{-0.7}$
                                                   &  8.3$^{+0.9}_{-0.6}$\\
         & $EW$ (eV)                               &  76 $\pm$ 6
                                                   &  81$^{+8}_{-7}$
                                                   &  79$^{+8}_{-6}$\\
Blue peak & $E$ (keV)                              & 6.98$^{+0.11}_{-0.10}$  
                                                   & 6.98 $\pm$ 0.09 
                                                   &  6.98 $\pm$ 0.09 \\ 
\multicolumn{2}{l}{$\beta$}                        & 2.1$^{+0.2}_{-0.6}$ 
                                                   & 2.1$^{+0.4}_{-0.3}$ 
                                                   & 2.1$^{+0.4}_{-0.3}$      \\
\multicolumn{2}{l}{$R_{\rm in}$}                   & $<$43 
                                                   & $<$32 
                                                   & $<$29   \\     
\multicolumn{2}{l}{$i$}                            & $<$10$\degr$    
                                                   & $<$12$\degr$    
                                                   & $<$10$\degr$     \\
Edge         & $E$ (keV)                           &  7.29$^{+0.25}_{-0.17}$ & ... & ... \\
             & optical depth $\tau$                &  0.034 $\pm$ 0.013 & ... & ...\\
Gaussian     & $E$ (keV)                          & ... &  7.68$^{+0.04}_{-0.03}$ & ... \\
             & $\sigma$                            & ... & $<$100 & ... \\
             & $\vert$ $I$ $\vert$ (10$^{-5}$ ph cm$^{-2}$ s$^{-1}$)   & ... & 1.0 $\pm$ 0.3 & ... \\
             & $\vert$$EW$$\vert$ (eV)             & ... & 13 $\pm$ 5 & ...\\      
Photoionized Abs. &  Col.\ dens.\ (10$^{22}$  cm$^{-2}$)      & ... & ... & 1.4$^{+1.0}_{-0.5}$  \\
                  &  log $\xi$ (erg cm s$^{-1}$)   & ... & ... & 3.73$^{+0.15}_{-0.13}$ \\ 
                  & z (rel. to systemic)           & ... & ... & --0.093$^{+0.006}_{-0.002}$ \\
\enddata
\tablecomments{Spectral parameters for best-fit models which attempt to
model the 7.68 keV absorption feature. The feature is too narrow
to be modeled by an edge, as demonstrated by Model 11.
However, modeling the feature with an inverted Gaussian (Model 12)
or using an {\sc xstar} model (Model 13) proved successful.
See text for additional details.}
\end{deluxetable}

%%%%%%%%%%%%%%%%%%%%%%%%%%%%%%%%%%%%%%%%%%%%%%%%%%%

\begin{deluxetable}{llc}
\tabletypesize{\footnotesize}
\tablewidth{6.5in}
\tablenum{8}
\tablecaption{Joint {\it RXTE}/EPIC-pn model fits \label{tab8}}
\tablehead{    \colhead{}&\colhead{}&\colhead{}   }
\startdata
\multicolumn{2}{l}{$\chi^2/dof$}                &     1874.75/1807 \\
\multicolumn{2}{l}{$\Gamma$ ({\it RXTE})}       &     1.822 $\pm$ 0.010  \\
\multicolumn{2}{l}{$\Gamma$ (pn)}               &     1.716$^{+0.013}_{-0.009}$    \\
\multicolumn{2}{l}{$R$}                         &     0.36 $\pm$ 0.04                \\
\multicolumn{2}{l}{$N_H$ (10$^{20}$ cm$^{-2}$)} &     $<$9        \\
Red peak & $E$ (keV)                               &  6.44 $\pm$ 0.01       \\
         & $I$ (10$^{-5}$ ph cm$^{-2}$ s$^{-1}$)   &  12.8 $\pm$ 1.0       \\
         & $EW$ (PCA; eV)                          &  149 $\pm$ 12    \\
         & $EW$ (pn; eV)                           &  126 $\pm$ 10   \\
Blue peak & $E$ (keV)                              &  6.91 $\pm$ 0.07 \\
\multicolumn{2}{l}{$\beta$}                        &  2.3 $\pm$ 0.1    \\
\multicolumn{2}{l}{$R_{\rm in}$}                   &  $<$7 $R_{\rm g}$             \\
\multicolumn{2}{l}{$i$}                            &  10 $\pm$ 7$\degr$          \\
Photoionized Abs. &  Col.\ dens.\ (10$^{22}$ cm$^{-2}$)      & 1.2$^{+0.3}_{-1.0}$  \\
                  &  log $\xi$ (erg cm s$^{-1}$)   & 3.99$^{+0.01}_{-0.33}$  \\
                  & z (rel. to systemic)           &  --0.093$^{+0.007}_{-0.004}$ \\
\enddata
\tablecomments{Spectral parameters for PCA, HEXTE and EPIC-pn fits
using the XSTAR model.
The photon indices for the PCA and HEXTE data were tied; that for
the EPIC-pn was left free. The inclination of the {\sc pexrav} component was
set at 10$\degr$, the best-fit inclination of the {\sc Laor} components. Errors on 
log$\xi$ pegged at the hard upper limit of 4.00.}
\end{deluxetable}

\begin{deluxetable}{llcc}
\tabletypesize{\footnotesize}
\tablewidth{6.5in}
\tablenum{9}
\tablecaption{Model 12 fits to each half of the EPIC-pn spectrum\label{tab9}}
\tablehead{
\colhead{} & \colhead{} & \colhead{First half} & \colhead{Second half}}
\startdata
\multicolumn{2}{l}{$F_{2-10}$ (erg cm$^{-2}$ s$^{-1}$)}                     & 9.0 $\times$ 10$^{-11}$ & 8.8 $\times$ 10$^{-11}$ \\
\multicolumn{2}{l}{$\chi^2/dof$}      & 1457.46/1492  & 1513.21/1490 \\
\multicolumn{2}{l}{$\Gamma$}          & 1.752$^{+0.019}_{-0.017}$ & 1.732$^{+0.021}_{-0.011}$  \\
\multicolumn{2}{l}{$N_H$ (10$^{20}$ cm$^{-2}$)}    & 15$^{+9}_{-10}$ & $<$14           \\                     
Red peak & $E$ (keV)                               &  6.46 $\pm$ 0.05        & 6.45${+0.01}_{-0.02}$ \\ 
         & $I$ (10$^{-5}$ ph cm$^{-2}$ s$^{-1}$)   &  8.2$^{+1.0}_{-0.9}$ & 8.8$^{+1.4}_{-1.1}$ \\
         & $EW$ (eV)                               &  78$^{+10}_{-9}$  & 86$^{+14}_{-11}$ \\ 
Blue peak & $E$ (keV)                              &  7.04$^{+0.10}_{-0.13}$ & 7.01 $\pm$ 0.12  \\ 
\multicolumn{2}{l}{$\beta$}                        & 3.3$^{+0.5}_{-2.0}$  & 2.1$^{+0.4}_{-0.7}$   \\
\multicolumn{2}{l}{$R_{\rm in}$}                   &  $<$70 &  $<$35              \\
\multicolumn{2}{l}{$i$}                            & $<$22$\degr$ & $<$12$\degr$ \\
Gaussian     & $E$ (keV)                           & 7.68 $\pm$ 0.06        & 7.69 $\pm$ 0.07        \\
             & $\sigma$                            & $<$150  & $<$200    \\
             & $\vert$ $I$ $\vert$ (10$^{-5}$ ph cm$^{-2}$ s$^{-1}$)   & 1.2$^{+0.7}_{-0.6}$ & 0.8 $\pm$ 0.5       \\ 
             & $\vert$$EW$$\vert$ (eV)             & 15$^{+9}_{-8}$ & 11 $\pm$ 7     \\
             & F-test to include                   & 99.8$\%$ & 95.8$\%$ \\ 
\enddata
\tablecomments{Best-fitting model 12, featuring dual-disklines for the Fe K emission peaks
and an inverted Gaussian for the 7.68 keV absorption feature, to the first and second halves
of the EPIC-pn data. All spectral parameters are consistent with remaining constant over timescales
$<$136 ksec. The final row shows the level of significance to include the inverted Gaussian
according to an F-test used in the ``standard'' manner (see text for details).}
\end{deluxetable}

%%%%%%%%%%%%%%%%%%%%%%%%%%%%%%%%%%%%%%%%%%%%%%%%%%%

\begin{deluxetable}{lcccc}
\tabletypesize{\footnotesize}
\tablewidth{6.5in}
\tablenum{10}
\tablecaption{F-test results to thaw $\Gamma$ or $I_{\rm Fe}$ in the time-resolved fits\label{tab10}}
\tablehead{
\colhead{Time} & \colhead{$\Gamma$} & \colhead{$\Gamma$} & \colhead{$I_{\rm Fe}$}  & \colhead{$I_{\rm Fe}$} \\
\colhead{Scale} & \colhead{F} & \colhead{Prob} & \colhead{F} & \colhead{Prob} }
\startdata
Short  &  0.15  & 0.99                     &   0.02  & 1.00 \\
Medium &  5.53  & 2.3 $\times$ 10$^{-12}$  &   1.51  & 7.8 $\times$ 10$^{-2}$ \\
Long   &  3.04  & 2.3 $\times$ 10$^{-5}$   &   2.28  & 1.9 $\times$ 10$^{-3}$ \\
\enddata
\tablecomments{Results of F-tests to determine which parameters it was significant to thaw in the
time-resolved fits. High values of the F-statistic and low-values of the probability
(of observing that value of F from a random set of data) indicate that the fits 
show improvement to thaw that parameter, suggesting statistically significant variations.}
\end{deluxetable}

%%%%%%%%%%%%%%%%%%%%%%%%%%%%%%%%%%%%%%%%%%%%%%%%%%%%%

\begin{deluxetable}{lccccccc}
\tabletypesize{\footnotesize}
\tablewidth{6.5in}
\tablenum{11}
\tablecaption{Mean spectral fit variability parameters and errors \label{tab11}}
\tablehead{
\colhead{Time} & \colhead{Mean $F_{2-10}$} & \colhead{Mean} & \colhead{Mean $I_{\rm Fe}$}  & \colhead{$\Gamma$} & \colhead{$\Gamma$} & \colhead{$I_{\rm Fe}$} & \colhead{$I_{\rm Fe}$} \\
\colhead{Scale} &\colhead{10$^{-11}$ ph cm$^{-2}$ s$^{-1}$} & \colhead{$\Gamma$} & \colhead{10$^{-5}$ ph cm$^{-2}$ s$^{-1}$} & \colhead{$r$} & \colhead{Prob} & \colhead{$r$} & \colhead{Prob}  }
\startdata
Short  &  8.91 $\pm$ 0.12 & 1.771 $\pm$ 0.027 &  8.22 $\pm$ 1.03 & 0.809 & 1.5 $\times$ 10$^{-2}$ & --0.099  & 0.82 \\
Medium & 13.29 $\pm$ 0.06 & 1.828 $\pm$ 0.033 & 20.47 $\pm$ 3.43 & 0.819 & 3.2 $\times$ 10$^{-5}$ &   0.272  & 0.27 \\
Long   & 15.62 $\pm$ 0.09 & 1.917 $\pm$ 0.067 & 16.60 $\pm$ 4.12 & 0.656 & 3.1 $\times$ 10$^{-3}$ & --0.059  & 0.82  \\
\enddata
\tablecomments{Mean values of $F_{2-10}$, $\Gamma$ and $I_{\rm Fe}$. See text for details on
error calculations. The Pearson correlation coefficients $r$ in Columns (4) and (6)
are for $\Gamma$ and $I_{\rm Fe}$, respectively, plotted against $F_{2-10}$, as shown in
Figure 11.  The corresponding null hypothesis probabilities (of achieving that value of $r$ from
a random set of data) are listed in Columns (5) and (7). See $\S$4.4 for notes relating PCA and pn
flux normalizations.}
\end{deluxetable}

%%%%%%%%%%%%%%%%%%%%%%%%%%%%%%%

\begin{deluxetable}{lcc}
\tabletypesize{\footnotesize}
\tablewidth{6.5in}
\tablenum{12}
\tablecaption{$F_{var}$ values\label{tab12}}
\tablehead{
\colhead{Time} & \colhead{Continuum} & \colhead{Line} \\
\colhead{Scale}& \colhead{$F_{\rm var}$ ($\%$)}    & \colhead{$F_{\rm var}$ ($\%$)} }
\startdata
Short  &    3.46 $\pm$ 0.48  &     Undef. \\
Medium &    9.34 $\pm$ 0.14  &   3.49 $\pm$ 13.99 \\ 
Long   &   12.99 $\pm$ 0.17  &   5.54 $\pm$ 19.42  \\
\enddata
\tablecomments{Undefined fractional variability measurements indicate a measured variance that is smaller
than that expected solely from measurement noise.}
\end{deluxetable}

%%%%%%%%%%%%%%%%%%%%%%%%%%%%%%%%%%

\clearpage

\begin{figure}
\epsscale{0.55}
\plotone{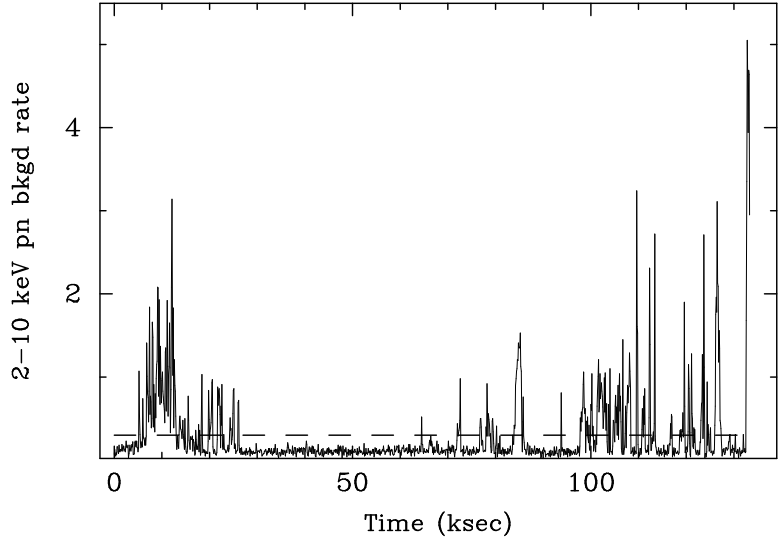}
%%%%\vspace{-6cm}
\caption{10--12 keV pn background light curve.
The dotted line at 0.3 count s$^{-1}$ corresponds to 
a count rate of $B$ + 2$\sigma$, 
where $B$ is the mean background rate and $\sigma$ is the 
standard deviation of the light curve.
We tested for background contamination by
filtering out data taken when the background
count rate exceeded this threshold.
However, we found no impact on the Fe K profile,
the focus of this paper, and used the unfiltered
data for maximum signal-to-noise ratio in the analysis.}
\end{figure}

%\clearpage

\begin{figure}
\epsscale{0.55}
\plotone{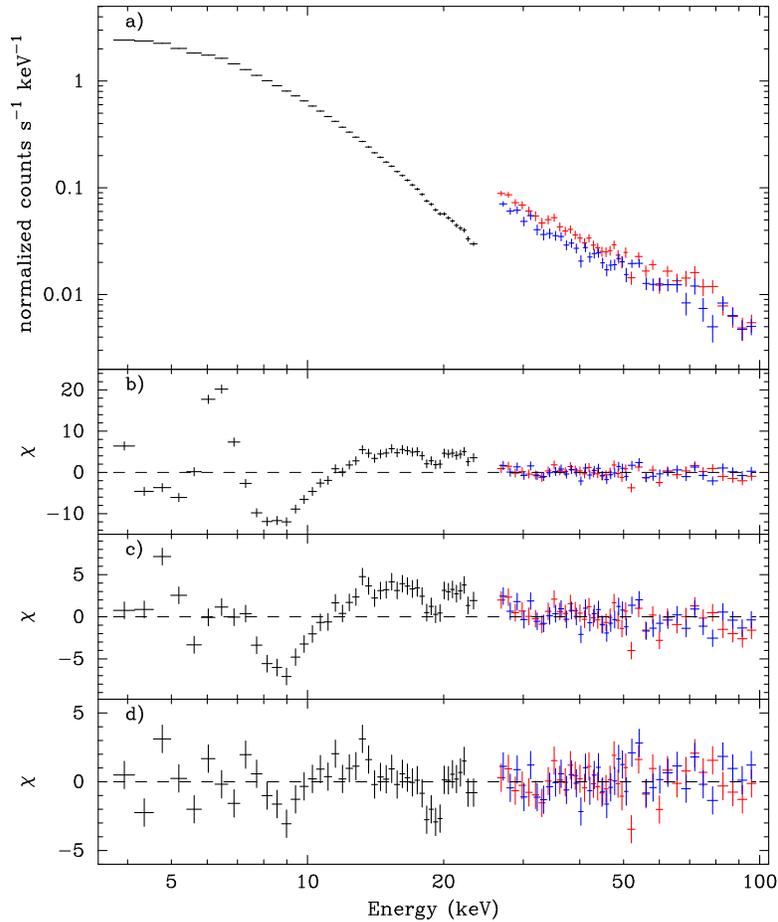}
%%%%%%\vspace{-1cm}
\caption{Panel a) shows the \xte\ PCA, HEXTE A cluster and HEXTE B cluster
spectra, in black, red and blue, respectively. 
Panel b) shows the data/model residuals when the data are fit with Model 1, an absorbed
power-law model.
Panel c) shows the residuals when a Gaussian is included to model Fe K emission
(Model 2). Panel d) shows the residuals when a {\sc pexrav}
reflection component is also included. 
For clarity in panels b), c), and d), residuals in $\chi$-space are plotted 
(typical uncertainties on the HEXTE data/model ratio points 
were $\pm$0.2 and so data/model ratios are not plotted).}
\end{figure}

\clearpage

\begin{figure}
\epsscale{0.70}
\plotone{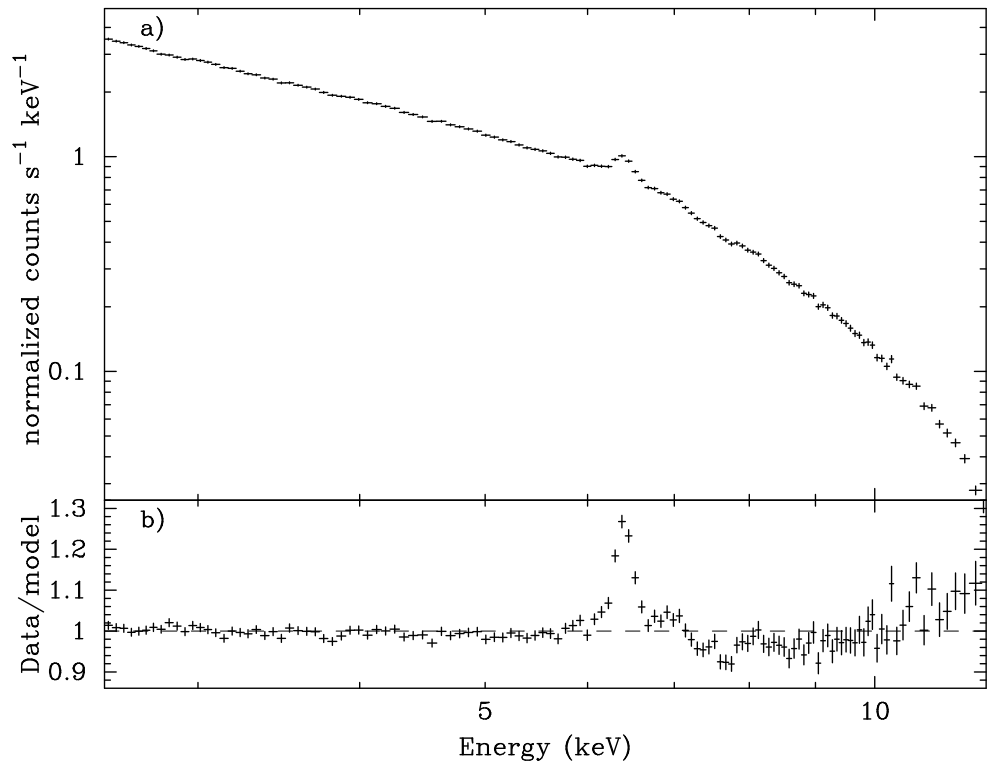}
%%%%%%%\vspace{-3cm}
\caption{Panel a) shows the 2.5--12 keV EPIC pn spectrum, with a binning factor of 15.
Panel b) shows the data/model residuals when a simple absorbed power-law
model is fit (Model 1). }
\end{figure}

\clearpage

\begin{figure}
\epsscale{0.50}
\plotone{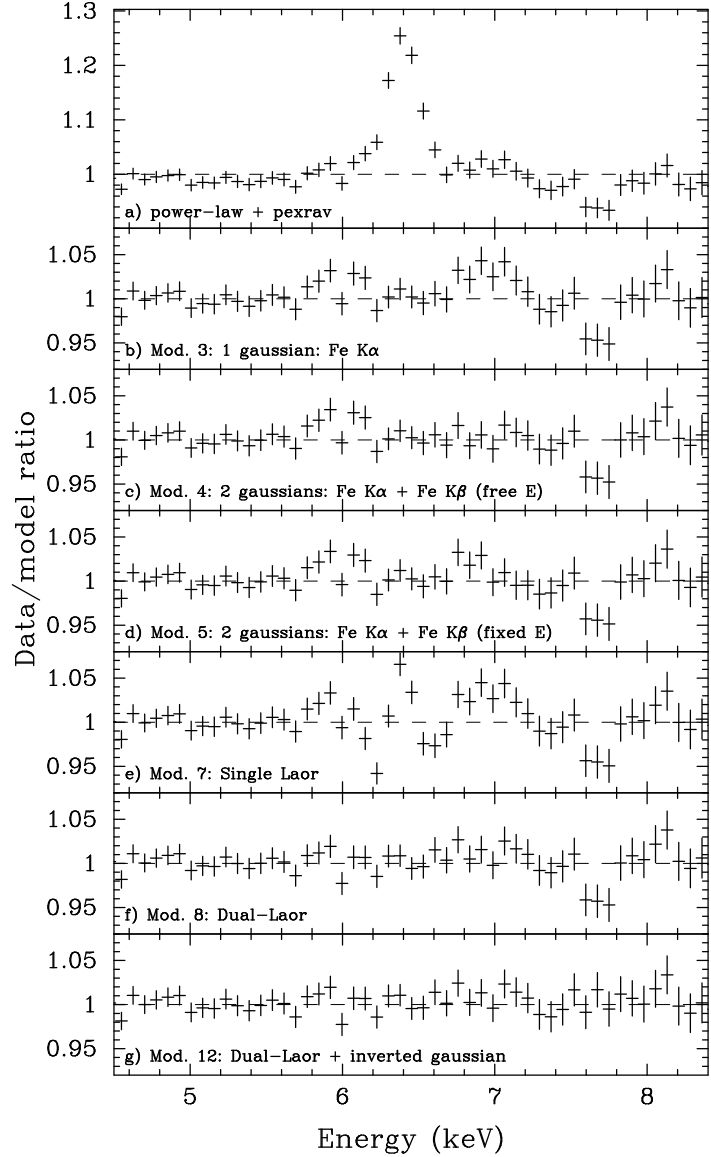}
%%%%%%%\vspace{-2cm}
\caption{\footnotesize Data/model residuals in the Fe K bandpass when various models are 
fit to the EPIC pn data. All plots here denote models that include 
a power-law and a {\sc pexrav} reflection component with $R$ fixed 
at --0.51, with all components modified by neutral absorption. Data 
plotted here have been rebinned by a factor of 15. Panel a) shows the residuals 
using just the power-law and {\sc pexrav} component (e.g., the 
$\sim$7.1 keV edge has been fit compared to the residuals in Figure 3b).
Panel b) shows the residuals to Model 3, which included a Gaussian to 
model Fe K$\alpha$ emission. Panels c) and d) show the residuals to 
Models 4 and 5, double Gaussian models for the Fe K$\alpha$ and 
K$\beta$ emission lines. The energy centroid for the K$\beta$ line 
was free in Model 4, but fixed at 7.06 keV in Model 5. Panel e) shows 
the residuals to Model 7, an attempt to model both peaks with a 
single {\sc Laor} diskline component; the large residuals 
suggest a poor fit to the data. Panel f) shows the residuals 
to Model 8, wherein each peak was modeled with a separate 
{\sc Laor} component. Note that the residuals near 5.8--6.1 keV and
6.7--6.9 keV are improved compared to Model 5, suggesting that 
the dual-{\sc Laor} model is fitting red wing emission unmodeled 
by the double-Gaussian model (see text for quantification of this 
statement). Note also that the absorption-like residuals near 7.68 keV
are unchanged in panels a)--f), suggesting that the absorption 
feature persists regardless of how the emission profile is modeled.
Panel g) shows the residuals to Model 12, which features 
dual-{\sc Laor} disklines to model the emission lines and a
narrow inverted Gaussian to model the 7.68 keV absorption feature.}
\end{figure}

\clearpage

\begin{figure}
\epsscale{0.50}
\plotone{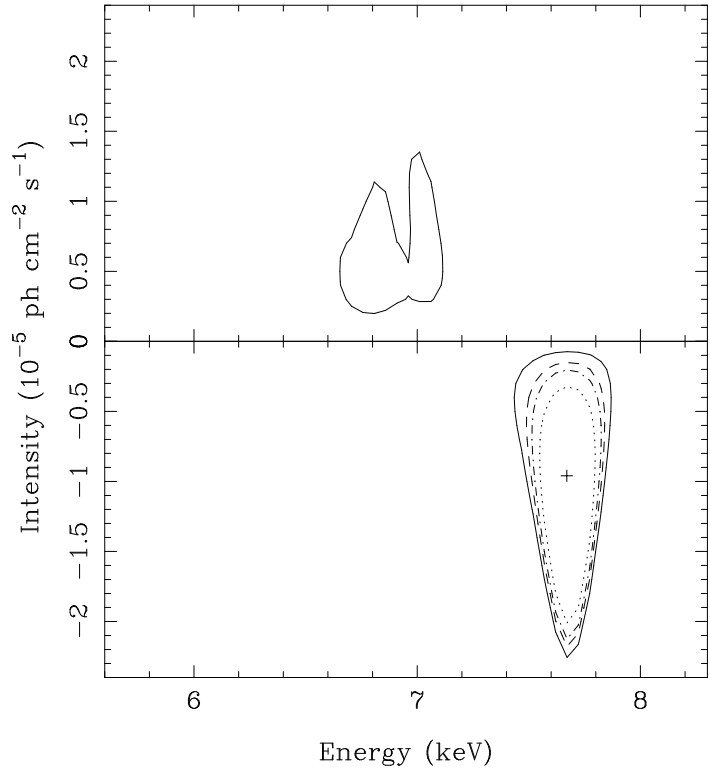}
\caption{Derived confidence contours of line intensity versus rest-frame 
energy when a narrow Gaussian is added in emission (upper panel) or absorption (lower panel)
to Model 8, the dual-{\sc Laor} diskline model. We stepped through
energy values in the range 5.60--8.30 keV in increments
of 0.05 keV. and flux range 0--3 $\times 10^{-5}$ ph cm$^{-2}$ s$^{-1}$
for emission (0 to --2.2 $\times 10^{-5}$ ph cm$^{-2}$ s$^{-1}$ for absorption)
in increments of $1 \times 10^{-6}$ ph cm$^{-2}$ s$^{-1}$.
The width $\sigma$ was kept fixed at 100 eV. Solid, dashed, dash-dotted, 
and dotted lines denote 68$\%$, 90$\%$, 95$\%$ and 99$\%$ confidence 
levels, respectively. There was no strong evidence from these plots for emission or
absorption at the \ion{Fe}{25} or \ion{Fe}{26} rest-frame energies. 
The upper panel shows there is no additional emission 
(at 90$\%$ confidence or greater) to be modeled. The lower panel 
shows a narrow absorption feature near 7.7 keV (at $>$99$\%$ confidence). 
The best-fitting parameters to Model 12 are 
marked with a cross. See text for details.}
\end{figure}

\begin{figure}
\epsscale{0.60}
\plotone{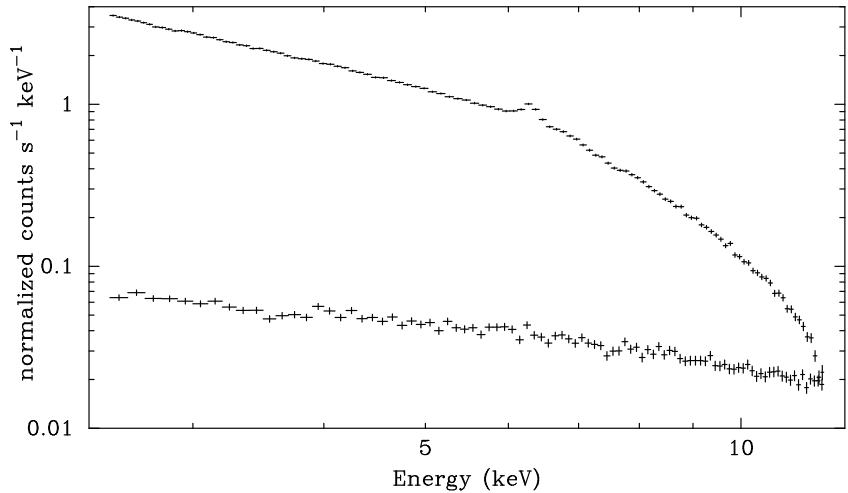}
%%%%\vspace{-5cm}
\caption{The pn source (top) and background (bottom) spectra, rebinned by a 
factor of 20, showing that the source is over an order of 
magnitude higher in counts s$^{-1}$ keV$^{-1}$ compared to the 
background in the Fe K bandpass.}
\end{figure}

\begin{figure}
\epsscale{0.75}
\plotone{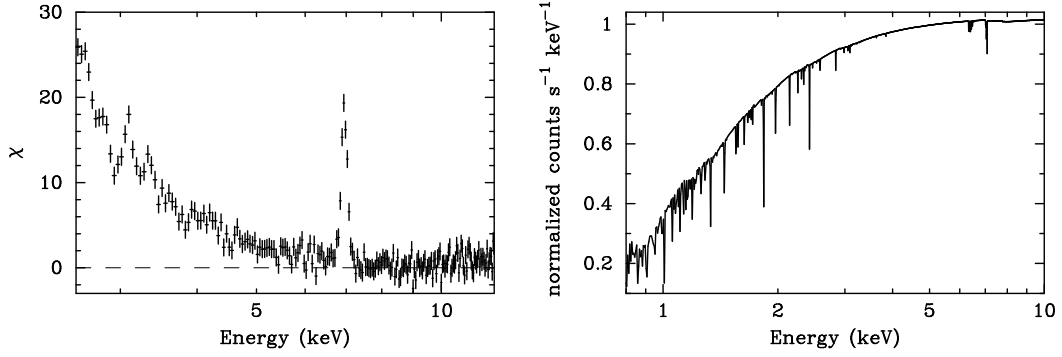}
\caption{ (left) -- A model with 
log$\xi$ = 2.0, a blueshift of 0.07$c$ relative to systemic,
and a column density 1.1 $\times$ 10$^{22}$ cm$^{-2}$
fits the 7.68 keV feature well. However, the model predicts
very strong spectral curvature below 4 keV,
as well as strong absorption at 6.9 keV due to blueshifted
K$\alpha$ absorption. See text for details.
(right) -- The effect     
of the total absorption from all the soft X-ray absorbers
in S05 on a simple power-law, normalized to $\sim$1 
at high energies, is shown. There is 
a negligible effect at energies above $\sim$4 keV.
The 7.68 keV absorption line is thus 
independently detected, and likely physically distinct, from
the soft X-ray absorbers. The small ($\sim$1 eV $EW$) absorption features
near the rest-frame energies for Fe K$\alpha$ and K$\beta$ emission
are not detectable, given the pn energy resolution
(data courtesy of K.\ Steenbrugge). }
\end{figure}

\begin{figure}
\epsscale{0.60}
\plotone{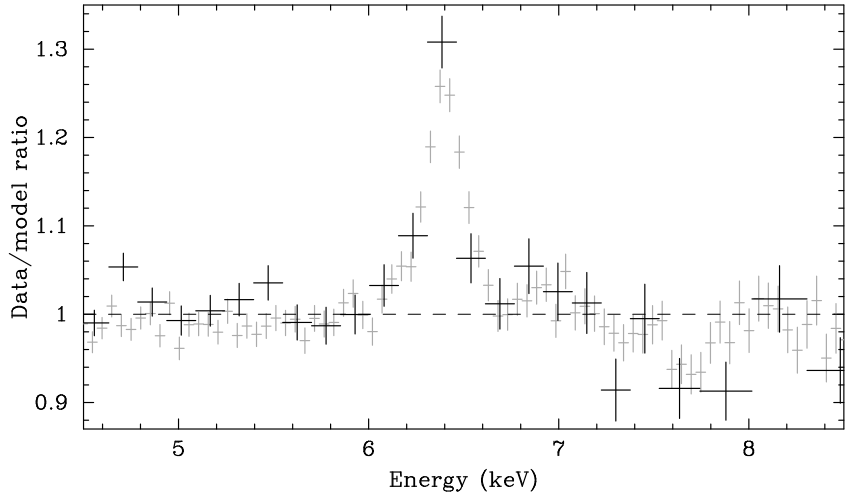}
%%%%\vspace{-5cm}
\caption{Data/model residuals to a simple absorbed power-law
(rebinned every 10 bins) for the MOS2 (black)
and pn (grey), showing that the MOS2 data are consistent
with the pn.}
\end{figure}

\begin{figure}
\epsscale{0.60}
\plotone{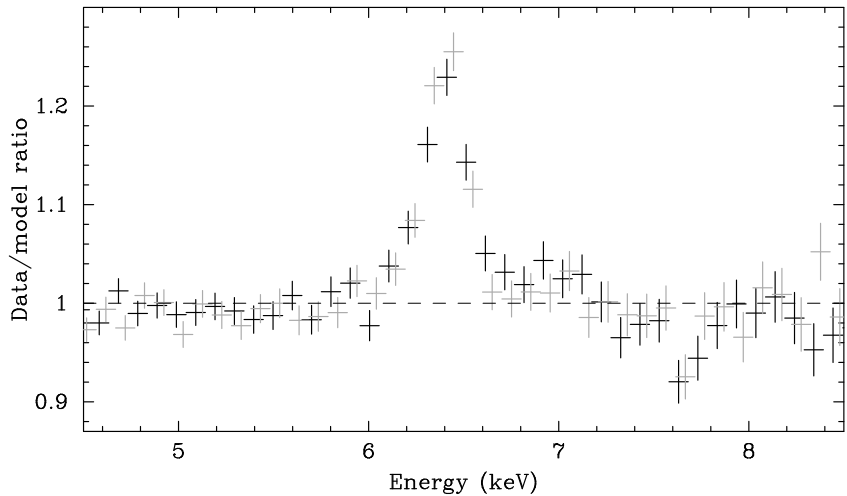}
%%%%%\vspace{-6cm}
\caption{Data/model residuals for the first half (black)
and second half (gray) of the pn data (rebinned every 20 bins) 
when fitting to a model consisting of a power-law
and a {\sc pexrav} reflection component, all modified by
neutral absorption. All derived spectral parameters are
consistent between the two halves.}
\end{figure}

\begin{figure}
\epsscale{0.80}
\plotone{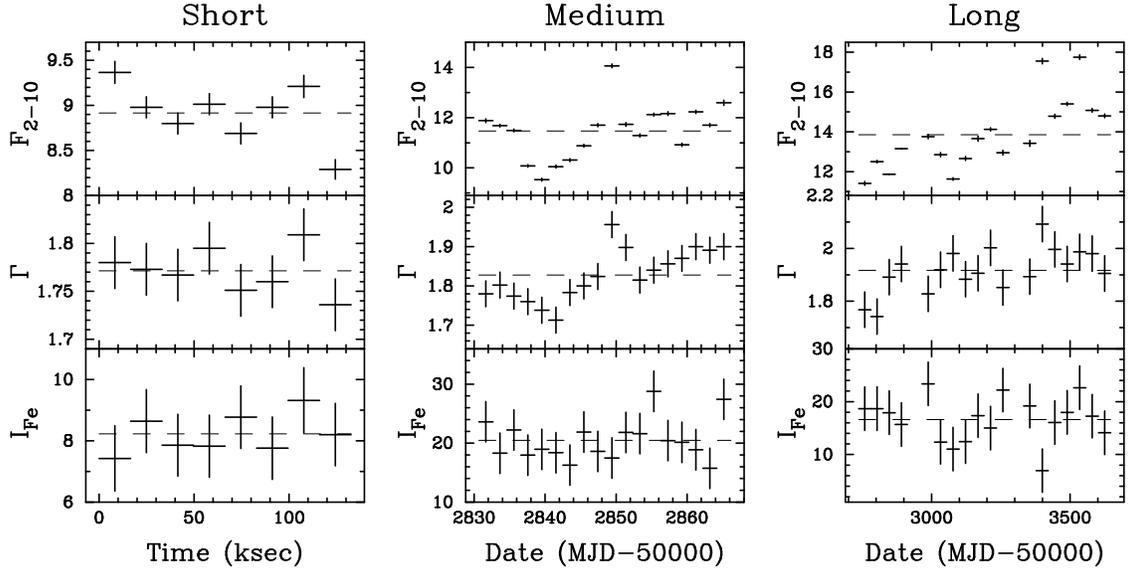}
%%%%%\vspace{-8cm}
\caption{Light curves for the 2--10 keV continuum flux $F_{2-10}$ (top),
photon index $\Gamma$ (middle), and Fe K flux $I_{\rm Fe}$ (bottom)
for all three time scales, derived from time-resolved spectroscopy.
$F_{2-10}$ is in units of 10$^{-11}$ erg cm$^{-2}$ s$^{-1}$, and
$I_{\rm Fe}$ is in units of 10$^{-5}$ ph cm$^{-2}$ s$^{-1}$.}
\end{figure}

\begin{figure}
\epsscale{0.80}
\plotone{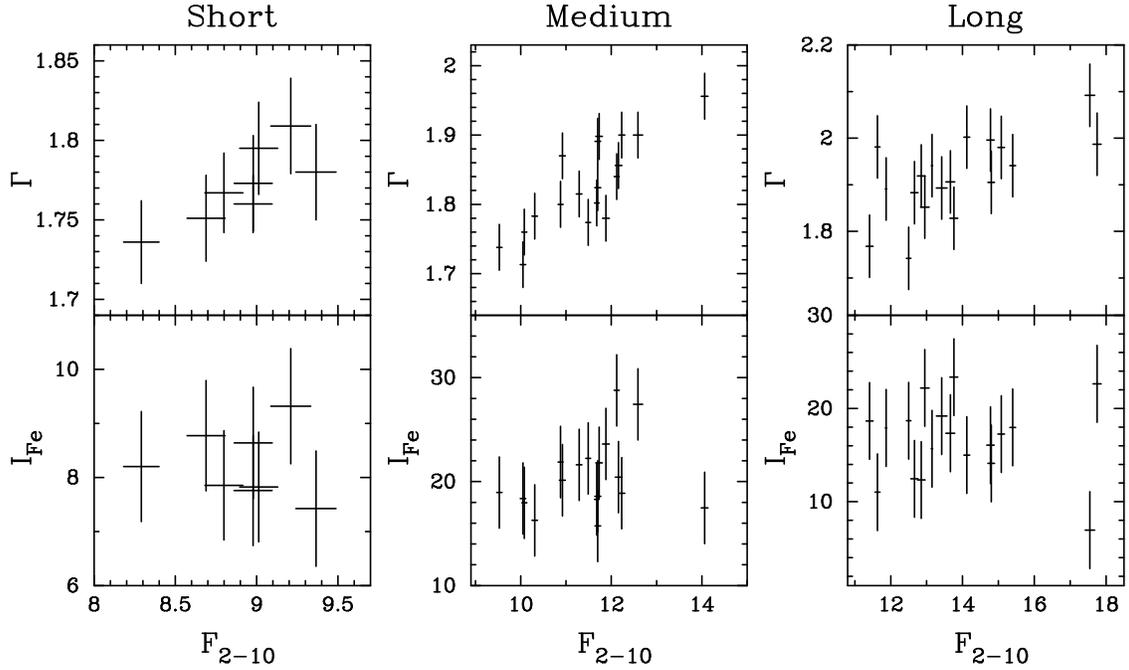}
%%%%%\vspace{-8cm}
\caption{Photon index $\Gamma$ (top) or Fe K flux $I_{\rm Fe}$ (bottom)
plotted against 2--10 keV continuum flux $F_{2-10}$
for all three time scales, derived from time-resolved spectroscopy.
$F_{2-10}$ is in units of 10$^{-11}$ erg cm$^{-2}$ s$^{-1}$, and
$I_{\rm Fe}$ is in units of 10$^{-5}$ ph cm$^{-2}$ s$^{-1}$.
These plots, along with Fig.\ 10 and Tables 10 and 11, suggest that 
temporal variations in $\Gamma$ are well-correlated
with those in $F_{2-10}$ on all three time scaled studied, but
there is no strong evidence for $I_{\rm Fe}$ to vary on any of
the time scales studied.}
\end{figure}

\end{document}